\documentclass[aps,prxquantum,twocolumn,superscriptaddress,10pt,floatfix]{revtex4-2}

\usepackage[T1]{fontenc}
\usepackage[utf8]{inputenc}
\usepackage{amsmath,amssymb,amsthm}
\usepackage{mathrsfs}
\usepackage{physics2}
\usephysicsmodule{ab,ab.braket}
\usepackage{derivative}
\usepackage{enumerate}
\usepackage{enumitem}
\usepackage{longtable}
\usepackage{multirow}
\usepackage{graphicx}
\usepackage{xcolor}
\usepackage{bm}
\usepackage{comment}
\usepackage{diagbox} 
\usepackage[colorlinks=true,allcolors=blue]{hyperref}

\DeclareMathOperator{\Arccos}{Arccos}
\DeclareMathOperator{\sgn}{sgn}

\DeclareMathOperator{\Sing}{Sing}

\newtheorem{thm}{Theorem}
\newtheorem{dfn}[thm]{Definition}
\newtheorem{lem}[thm]{Lemma}
\newtheorem{prop}[thm]{Proposition}
\newtheorem{cor}[thm]{Corollary}

\begin{document}

\title{Two-dimensional quantum central limit theorem by quantum walks}

\author{Keisuke Asahara}
\affiliation{Education and Research Center for Mathematical and Data Science, Hokkaido University, Kita 12, Nishi 7, Kita-ku, Sapporo, Hokkaido, 060-0812, Japan}
\email{keisuke-asahara@mdsc.hokudai.ac.jp}

\author{Daiju Funakawa}
\affiliation{Department of Electronics and Information Engineering, Hokkai-Gakuen University, Sapporo, Hokkaido 062-8605, Japan}
\email{funakawa@hgu.jp}

\author{Motoki Seki}
\affiliation{Department of Mathematics, Faculty of Science, Hokkaido University, Sapporo, 060-0810, Japan}
\email{marmot.motoki@gmail.com}

\author{Akito Suzuki}
\affiliation{Department of Production Systems Engineering and Sciences, Komatsu University, Awazu Campus, Nu 1-3 Shicho-machi, Komatsu, Ishikawa 923-8511, Japan}
\email{akito.suzuki@komatsu-u.ac.jp}

\begin{abstract}
The weak limit theorem (WLT), the quantum analogue of the central limit theorem, is foundational to quantum walk (QW) theory. 
Unlike the universal Gaussian limit of classical walks, 
deriving analytical forms of the limiting probability density function (PDF) in higher dimensions has remained a challenge 
since the 1D Konno distribution was established. 
Previous explicit PDFs for 2D models were limited to specific cases whose fundamental nature was unclear.

This paper resolves this long-standing gap by introducing the notion of maximal speed $v_{\mathrm{max}}$ as a critical parameter. 
We demonstrate that all previous 2D solutions correspond to a degenerate regime where $v_{\mathrm{max}} = 1$. 
We then present the first exact analytical representation of the limiting PDF for the physically richer, unexplored regime $v_{\mathrm{max}} < 1$ of a general class of 2D two-state QWs.

Our result reveals 2D Konno functions that govern these dynamics. 
We establish these as the proper 2D generalization of the 1D Konno distribution by demonstrating their convergence to the 1D form in the appropriate limit. Furthermore, our derivation, based on spectral analysis of the group velocity map, analytically resolves the singular asymptotic structure: we explicitly determine the caustics loci where the PDF diverges and prove they define the boundaries of the distribution's support. By also providing a closed-form expression for the weight functions, this work offers a complete description of the 2D WLT.

\end{abstract}

\maketitle

\section{Introduction}
The \emph{central limit theorem} (CLT) for classical random walks stands as a cornerstone of stochastic modeling, underpinning their vast applications 
\cite{Kac1947,
feller1968introduction, weiss1994aspects, motwani1995randomized,
grimmett2020probability,
Yan2023Diffusionmodels}. 
Inspired by this classical success, and fueled by the rise of quantum computation
\cite{feynman1982simulating, Benioff1980TheCA, DeutschJozsa1992, Grover1996, Lloyd1996UniversalQuantumSimulators, Shor1997, HornGottlib2001,
Alan2005simulatedQCofMolecularEnergies,
Perdomo2008, Kassal2009}, 
\emph{quantum walks} (QWs) were introduced as their quantum counterpart \cite{Aharanov1993PhysRevA.48.1687, FarhiGutmann1998PhysRevA.58.915, NayakAshvi2000QuantumwalkontheLine, AmbainisBachNayakVishwanathWatrous2001OnedimensionalQuantumwalks}. 
QWs have proven to be a powerful resource, demonstrating capabilities for 
\emph{universal quantum computation}
\cite{Childs2009UniversalComputationbyQWPhysRevLett.102.180501,
LovettCooperEverittTreversKendon2010PhysRevA.81.042330,
ChildsGossetWebb2013UniversalComputationbyMultiQWdoi:10.1126/science.1229957}
and offering new paradigms for quantum algorithms
\cite{Ambainis2003, Childs2003, Szegedy, Varsamisetal2023DNAassemblyquantumalgorithms}
and the simulation of physical phenomena
\cite{BialynickiBirula1994PhysRevD.49.6920, Meyer1996Quantumcellularautomata,
SucciBenzi1993LatticeBoltzmannEquation, BoghosianTaylor1996SimulatingQuantumMechanics,
MohseniRebentrostLloydAspuruGuzik2008,
Kitagawa2012, PanahiyanFritzsche2019}.

Given the foundational role of the CLT, establishing its quantum analogue---the \emph{weak limit theorem} (WLT)---
has been a central goal since Konno's pioneering 1D analyses
\cite{konnoQuantumRandomWalks2002, konnoNewTypeLimit2005},
which initiated an extensive line of inquiry
\cite{grimmettWeakLimitsQuantum2004,
inuiOnedimensionalThreestateQuantum2005,
Konno2005PhysRevE.72.026113,
miyazakiWignerFormulaRotation2007,
segawaLimitTheoremsQuantum2008,
konnoQuantumWalks2008a,
liuOnedimensionalQuantumRandom2009,
chisakiLimitTheoremsDiscreteTime2009,
konnoOnedimensionalDiscretetimeQuantum2009a,
konnoLimitTheoremsQuantum2010,
machidaLimitTheoremTimeDependent2010,
machidaLimitTheoremsLocalization2011,
liuAsymptoticDistributionsQuantum2012,
chisakiLimitTheoremsDiscretetime2012,
wojcikTrappingParticleQuantum2012,
konnoLimitTheoremsOpen2013,
liuWeakLimitsQuantum2013,
konnoLimitMeasuresInhomogeneous2013,
Machida2013realization_of_the_probability_laws,
tateAlgebraicStructureOnedimensional2013,
shikanoDiscreteTimeQuantum2013,
endoStationaryMeasureSpaceinhomogeneous2014,
endoOnedimensionalHadamardWalk2014a,
konnoUniformMeasureDiscretetime2014,
falknerWeakLimitThreestate2014,
konnoNonuniformStationaryMeasure2015,
suzukiAsymptoticVelocityPositiondependent2016,
endoWeakLimitTheorem2016,
richardQuantumWalksAnisotropic2018a,
richardQuantumWalksAnisotropic2018b,
maedaWeakLimitTheorem2018a,
FudaFunakawaSuzuki2019,
Wada2019,
mandalLimitTheoremsLocalization2022}.
However, the analogy is not straightforward. The WLT exhibits a rich, non-trivial structure, departing fundamentally from the universal Gaussian limit of the CLT. While this structure is now well-understood in 1D through the \emph{Konno distribution}, 
its generalization to higher dimensions has proven far more challenging.

Despite this two-decade effort, progress in higher dimensions remained limited;
while explicit PDFs were derived for some specific 2D models, their fundamental nature and relation to the 1D case were unclear \cite{watabeLimitDistributionsTwodimensional2008,
difrancoAlternateTwodimensionalQuantum2011}.

A key contribution of this paper is to demonstrate, 
by introducing a notion of maximal speed $v_{\mathrm{max}}$, 
that all these previous 2D solutions correspond 
to a degenerate regime where $v_{\mathrm{max}} = 1$.
Consequently, the physically richer, non-trivial regime 
$v_{\mathrm{max}} < 1$---which constitutes the \emph{proper} 2D generalization of the 1D Konno distribution---had remained an \emph{entirely open problem} since the original 1D work.

This paper directly addresses this \emph{long-standing gap}. 
We present the first exact analytical representation of the limiting PDF for a general class of 2D two-state QWs (2D2SQWs), resolving their asymptotic behavior in the non-trivial regime 
$v_{\mathrm{max}} < 1$ and providing a complete description of the limit distribution.

\subsection{Weak Limit Theorems}
\label{subsec:WLT}
To fully appreciate the significance of this 2D generalization, we begin by reviewing the foundational 1D WLT.
This line of inquiry, which seeks a quantum analogue of the CLT, 
was first pursued by Konno
\cite{konnoQuantumRandomWalks2002,
konno2005a_new_type_of_limit_theorem}
as a WLT
for a 1D2SQW,
which states that the average displacement of the position $X_t$
for a walker during $t$ steps
converges \emph{weakly} to
the asymptotic velocity $V$ as $t \to \infty$.

The long-time asymptotic behavior of QWs can be rigorously separated into three types
by introducing the \emph{maximal speed},
$v_{\mathrm{max}} \in [0,1]$,
derived from the group velocity $v = d\omega / dk$
(defined by the dispersion relation $\omega$)
\cite{Meyer1996QMgas1, DavidAMeyer_1998QMgas2, Strauch2006RelativisticQWsPhysRevA.73.05430}.
This parameterization allows for 
a clear summary of the 1D behaviors
in TABLE~\ref{tab:1}
\cite{konno2005a_new_type_of_limit_theorem,
venegas-andracaQuantumWalksComprehensive2012}.

We first explain the cases of (a) and (b),
which exhibit exceptional phenomena
\cite{suzukiAsymptoticVelocityPositiondependent2016,
richardQuantumWalksAnisotropic2018a,
richardQuantumWalksAnisotropic2018b}.

(a) \emph{Bound in a finite region.}
In the case of $v_{\mathrm{max}}=0$,
the asymptotic velocity is zero with probability one,
i.e., $P(V=0) = 1$.
This phenomenon,
sometimes referred to as \emph{Localization} \cite{mackayQuantumWalksHigher2002, inuiOnedimensionalThreestateQuantum2005},
wcomes from the evolution operator having only eigenvalues,
and thus the walker can only take a superposition of bound states,
which means that the walker can hardly escape a sufficiently large finite region (see \cite[Theorem 8.1]{Thaller1992Dirac} for instance).

(b) \emph{Trivial ballistic motion.}
The case of $v_{\mathrm{max}} = 1$ \emph{only} allows the walker to move in a single direction. 
If the walker starts to move to the left (or right),
the walker always moves to the left (or right).
This phenomenon comes from the No-Go lemma \cite{Meyer1996Quantumcellularautomata} which states
in our setting that
all scalar homogeneous unitary QWs are trivial,
that is, the walker can move in a single direction.
In fact, in the case of $v_{\mathrm{max}} = 1$,
the evolution operator can be represented as
the direct sum of two homogeneous unitary QWs
without any inner degree of freedom.

(c) \emph{Linear spreading.}
Among these cases listed in TABLE \ref{tab:1}, 
the parameter regime $v_{\mathrm{max}} \in (0,1)$
reveals the typical behavior of the QW.
The asymptotic velocity $V$ in this regime follows
the \emph{Konno distribution}:
\begin{align}
    P(V \leq u ) 
    & = \lim_{t \to \infty}P\left( \frac{X_t}{t} \leq u \right) \notag \\
    \label{Konnodist}
    & = \int_{-\infty}^u w(v) f_{\mathrm{K}}(v; v_{\mathrm{max}})dv,
\end{align} 
where the function $f_{\mathrm{K}}$ is the Konno function,
defined as
\begin{equation}
    \label{KonF1}
    f_{\mathrm{K}}(v; r)
    :=\dfrac{\sqrt{1 - r^2}}{\pi(1 - v^2)\sqrt{r^2 - v^2}}I_{(-r, r)}(v)
\end{equation}
with $r \in (0,1)$ being a parameter and $I_S$ is the indicator function of a set $S$. 
Crucially, while the maximal speed $v_{\mathrm{max}}$
and the Konno function $f_{\mathrm{K}}$
are independent of the initial state,
the function $w$ carries all the information
about the initial state in the distribution \eqref{Konnodist}.
Assuming that the walker starts at the origin
with the initial qubit state
$\alpha \ket|0> + \beta \ket|1>$,
we can interpret the original result by Konno
\cite{konnoQuantumRandomWalks2002,
konno2005a_new_type_of_limit_theorem}
in terms of effective mass
$m_{\mathrm{eff}}= \left(d^2\omega / dk^2|_{k = 0}\right)^{-1}$
\cite{Strauch2006RelativisticQWsPhysRevA.73.05430}
as
\begin{equation} \label{1Dw(v)startfromorigin}
    w(v) = 1-
    \left[
        |\alpha|^2 - |\beta|^2 + \left(\bar{\alpha}\beta + \bar{\beta}\alpha \right) m_{\mathrm{eff}}
    \right] v. 
\end{equation}
For the initial state of the form
$\psi \otimes (\alpha \ket|0> + \beta \ket|1>)$,
Machida \cite{Machida2013realization_of_the_probability_laws}
constructed interesting examples of explicit expressions for $w(v)$
by choosing functions $\psi$ appropriately.

\begin{table}[bt]
    \begin{tabular}{|l||c|c|} \hline 
        Maximal speed & Distribution of $V$ & Asymptotic behavior \\ \hline 
        (a) $v_{\mathrm{max}} = 0$ & $\delta_0$ & Bound in a region \\ \hline
        (b) $v_{\mathrm{max}}=1$ & $C_{+1} \delta_{+1} + C_{-1} \delta_{-1}$ & Trivial ballistic motion \\\hline
        (c) $v_{\mathrm{max}} \in (0,1)$ & $w(v) f_{\mathrm{K}}(v; v_{\mathrm{max}})$ & Linear spreading \\ \hline
    \end{tabular}
    \caption{Distribution of $V$ and corresponding asymptotic behaviors of 1D2SQW.
    Here, $\delta_a$ denotes the Dirac measure: $\delta_a(B)=1$ if $a \in B$; $0$ otherwise.}
    \label{tab:1}
\end{table}

However, for a general initial state $\Psi_0$,
determining $w(v)$ requires knowledge of the dynamics of the QW,
since the probability density function (PDF)
of the asymptotic velocity $V$
is given by the spectral measure \cite{ReedSimonI1980},
formally written as
$|\braket < v | \Psi_0>|^2 dv$,
of the velocity operator $\hat{v}$
\cite{grimmettWeakLimitsQuantum2004,
suzukiAsymptoticVelocityPositiondependent2016,
richardQuantumWalksAnisotropic2018a}
associated with $\Psi_0$.
The 1D2SQW consists of two wave packets
corresponding to the two modes of frequencies $\pm \omega$,
which inevitably makes the group velocity $v$
a two-valued function $v_{\pm}(k)$ of the wave number $k \in \mathbb{T}$.
Since each $v_{\pm}$ serves as a covering projection
from $\mathbb{T}$ to the velocity domain $|v|\leq v_{\mathrm{max}}$,
the wave number space must be partitioned
into injective regions $\mathbb{T} = \mathbb{T}_0 \cup \mathbb{T}_1$
to properly define the single-valued inverse functions
$k = k_{\star, m}(v) := v_{\star}|_{\mathbb{T}_m}^{-1}(v)$
($\star=\pm, \ m=0, 1$).
The rigorous application of the spectral analysis establishes
the following closed expression for $w(v)$
(see \cite{richardQuantumWalksAnisotropic2018b} for details):
\begin{equation} \label{wv}
    w(v) = \dfrac{1}{2} \sum_{\substack{\star = \pm \\ m = 0, 1}} 
    P_{\star}\bm{(}k_{\star,m}(v)\bm{)},
\end{equation}
where $P_{\star}(k)$ is a function calculated by 
the initial state.
The derivatives of the inverse functions $k_{\star, m}$
derive the Konno function \eqref{KonF1} up to a constant factor.

\subsection{Finding 2D Konno Function}
As mentioned, the generalization of the 1D Konno distribution to higher dimensions has been a long-standing challenge.
A critical distinction must be made here: 
while the WLT, which only ensures the \emph{existence} of the limit, has been proven for broad classes of multi-dimensional QWs \cite{grimmettWeakLimitsQuantum2004, Sako2021}, 
the \emph{explicit analytical forms} of their limiting PDFs have remained elusive.

To date, explicit analytical expressions for the limiting PDF have been derived only for the specific 2DQW models analyzed by
Watabe et al.~\cite{watabeLimitDistributionsTwodimensional2008}
and Di Franco et al.~\cite{difrancoAlternateTwodimensionalQuantum2011}.
They derived the following 2D function $f({\bm v})$
for these models:
\begin{equation}
\label{KonF2}
f({\bm v}) 
= \dfrac{1}{\pi^2 (1 - v_1^2)(1 - v_2^2)}I_{E}({\bm v}),
\end{equation}
where $E$ is an elliptic region given by
\begin{equation} \label{ellipse}
    E = \left \{
        {\bm v} = (v_1, v_2) \;\middle| \; 
        \dfrac{(v_1 + v_2)^2}{4a}  + \dfrac{(v_1 - v_2)^2}{4b} \leq 1 
    \right\}
\end{equation}
with a pair $(a, b)$ of constants satisfying $a + b = 1$.
As will be proved later, 
taking the 1D limit of Eq.~\eqref{KonF2}
results in the trivial ballistic behavior of $v_{\mathrm{max}} = 1$,
rather than the linear spreading described by the Konno function $f_{\mathrm{K}}$ for $v_{\mathrm{max}} \in (0, 1)$. 
Moreover, the accompanying challenge of analytically determining the initial-state-dependent function $w$ for general initial states remains unaddressed.

This task is formidable,
as the complexity of the spectral structure of
the 2D velocity operator---%
which involves summing over multiple inverse velocity branches
(four branches in the 1D case \cite{richardQuantumWalksAnisotropic2018b})---%
grows with dimension.
Unlike the 1D case where the velocity map $v_{\pm}(k)$ is a relatively simple covering, the 2D group velocity map $\bm{v}_{\pm}(\bm{k})$ is a highly \emph{non-injective} 2D-to-2D transformation. Deriving $w$ requires a complete spectral analysis, which hinges on inverting this map. This inversion is critically complicated by the presence of \emph{caustics}---loci in the velocity space corresponding to points $\bm{k}$ where the Jacobian of the velocity map vanishes (i.e., the effective mass tensor diverges \cite{ahlbrechtAsymptoticEvolutionQuantum2011}). These caustics manifest as divergences in the limiting PDF and, as suggested by Ahlbrecht et al.~\cite{ahlbrechtAsymptoticEvolutionQuantum2011} via numerical analysis, define the boundaries of the PDF's support. However, an analytical resolution of this \emph{singular structure} for a general 2DQW has remained elusive.

This critical omission, combined with the disconnection
between $f$ and $f_{\mathrm{K}}$,
raises fundamental questions:
\begin{enumerate}[leftmargin=2em]
\item \emph{What is a proper 2D generalization of the Konno function $f_{\mathrm{K}}$?}
\item \emph{What is a closed form of the weight function $w$?}
\item \emph{Can the singular asymptotic structure (the caustics and the PDF support) be determined analytically?}
\end{enumerate}

This work addresses these questions by first identifying the parameter constraint $a+b=1$
of the known 2DQW models 
\cite{watabeLimitDistributionsTwodimensional2008,
difrancoAlternateTwodimensionalQuantum2011} 
with the maximal speed, $v_{\rm max} = a+b$.
This identification reveals that previous analytical studies were confined to the $v_{\rm max}=1$ regime. 
As we will prove, this regime, 
when taken to the proper 1D limit, 
degenerates into the trivial ballistic motion of a 1D QW, in sharp contrast to the linear spreading regime described by the Konno distribution
(See TABLE~\ref{tab:1} (b) and (c)).

We therefore address these fundamental questions
by providing the first analytical solution for the unexplored, 
non-trivial regime $v_{\rm max} = a + b < 1$. 
Our main contributions are threefold,
corresponding directly to the questions above:

\begin{enumerate}[leftmargin=2em]
\item We obtain the exact representation 
of the \emph{2D Konno functions} $f_\pm$
for unexplored regime $v_{\rm max} < 1$,
addressing \emph{question 1}.
We rigorously establish these 
as the \emph{proper} 2D generalization of the Konno function
by demonstrating their convergence back 
to the 1D Konno function $f_{\rm K}$ in the appropriate 1D limit, 
whereas the known PDF \eqref{KonF1} degenerates to the trivial ballistic model.
\item We provide a complete description of the limit distribution by deriving the closed-form expression 
for the \emph{weight function} $w(v)$ for both regimes $v_{\rm max} \le 1$, 
thereby resolving \emph{question 2}. 
\item Our thorough spectral analysis provides the first complete analytical resolution of the 2DQW's \emph{singular asymptotic structure},
addressing \emph{question 3}.
We explicitly determine the \emph{caustics} loci where the Jacobian vanishes and the PDF diverges,
and prove they define the boundaries of the PDF's support. 
This analytically generalizes what was previously suggested numerically \cite{ahlbrechtAsymptoticEvolutionQuantum2011}.
\end{enumerate}

The remainder of this paper is organized as follows:
In Sec.~\ref{sec:models}, we introduce our 2DQW model and
characterize it with the pair of parameters $(a, b)$,
which defines the maximal speed $v_{\mathrm{max}} = a + b$.
Sec.~\ref{sec:main_theorem} presents our main results,
deriving the limiting PDFs for both parameter regimes:
For $v_{\mathrm{max}} \in (0, 1)$,
we derive the 2D Konno functions $f_{\pm}$ (Theorem~\ref{thm:WLT}).
For $v_{\mathrm{max}} = 1$,
we recover the known function $f$ (Theorem~\ref{thm:WLT'}).
Here, we also provides the rigorous 1D limit theorems
for the entire distribution. We prove that $v_{\rm max}<1$ regime
properly converges to the 1D Konno distribution (Theorem~\ref{thm:btozero}), while the $v_{\rm max}=1$ regime
degenerates to trivial ballistic motion (Theorem~\ref{thm:btozero1}).
These theorems establish \emph{contribution 1}.
Sec.~\ref{sec:discussion} validates and discusses 
the implications of these results.
We compare our findings with previous studies.
We then analyze the convergence \emph{mechanism} 
of the 2D functions $f_\pm$ themselves in Sec.~\ref{subsec:interpretation},
detailing how their support deforms (FIG.~\ref{fig:domain_def})
to recover the 1D Konno function $f_{\rm K}$.
Sec.~\ref{subsec:singularasymptotics}
discusses the physical implications of 
our analytical resolution of the singular asymptotics.
Sec.~\ref{sec:proofs} sketches the proofs for the main theorems.
This section details technical achievement of our work:
the spectral analysis of the velocity map, the handling of the Jacobian, and the identification of caustics (Theorem~\ref{thm:Jacobinv}),
which completes \emph{contribution 3};
and the final derivation of the closed-form weight function $w(v)$ (\eqref{eq:w_bullet} in Sec.~\ref{sec:DerivationPDF}),
which achieves \emph{contribution 2}.
Sec.~\ref{sec:conclusion} concludes this paper.
The appendices contain detailed proofs of propositions
and additional supporting results.

\section{Models} \label{sec:models}
This paper investigates a \emph{homogeneous} 2D2SQW,
a model initially defined by Di Franco et al.~\cite{diFranco_Mimicking_probability_dist_2dGW},
and referred to as the alternate QW (AQW)~\cite{diFranco_Mimicking_probability_dist_2dGW}. 
This model was originally introduced to realize 2D four-state QWs (2D4SQWs) \cite{mackayQuantumWalksHigher2002} such as the Grover walk \cite{inuiLocalizationTwodimensionalQuantum2004a} with fewer experimental resources. 
We treat a slightly generalized version of the 2D2SQW model.
As will be shown, this modification provides rich phenomena for the asymptotic behavior of the walker,
enabling the explicit derivation of the 2D Konno function.

\subsection{Probability and Dynamics}

\begin{figure}
	\includegraphics[width=8.6cm]{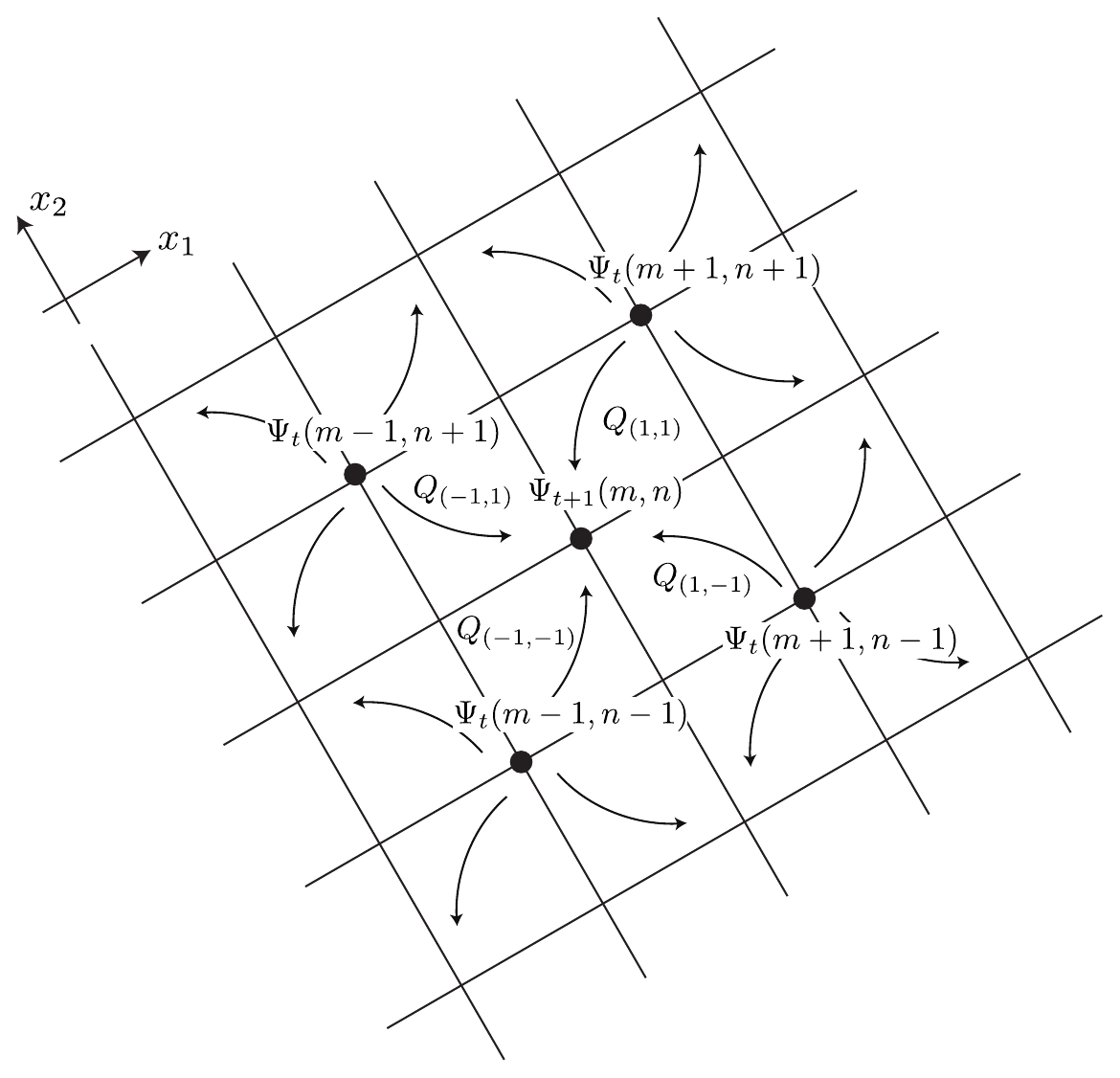}
	\caption{The walker in $\mathbb{Z}^2$ that follows \eqref{state_evol}.} \label{fig:2D2SQW}
\end{figure}

We consider a walker moving on the 2D square lattice
$\mathbb{Z}^2 = \{\bm{x} = (x_1, x_2) \mid x_1, x_2 \in \mathbb{Z}\}$.
The position $\bm{X}_t$ of the walker at time $t = 0, 1, 2, \dots$ follows the probability distribution:
\begin{equation}
\label{distXt}
    P(\bm{X}_t = \bm{x}) =  \Psi_t^\dagger(\bm{x}) \Psi_t (\bm{x}), \quad \bm{x} \in \mathbb{Z}^2,
\end{equation}
where the $\mathbb{C}^2$-valued function
$\Psi_t(\bm{x}) = \begin{pmatrix} \Psi_t^{(+)}(\bm{x}) \\ \Psi_t^{(-)}(\bm{x}) \end{pmatrix}$
represents the state of the walker
with each component $\Psi_t^{(\pm)}(\bm{x})$ denoting
the walker's inner degrees of freedom.
The Hilbert space of states of the walker 
is thus the set of
$\mathcal{H} := \ell^2(\mathbb{Z}^2; \mathbb{C}^2) \simeq \ell^2(\mathbb{Z}^2) \otimes \mathbb{C}^2$,
where $\ell^2(\mathbb{Z}^2)$ is the Hilbert space of 
square summable functions from $\mathbb{Z}^2$ to $\mathbb{C}$.
The state evolution of the 2D2SQW is then defined by
the recurrence relation:
\begin{equation}
\label{state_evol}
    \Psi_{t+1}(\bm{x}) = \sum_{\bm{s} \in \{(s_1,s_2) \mid s_j =-1,+1\}}
    Q_{\bm{s}} \Psi_t(\bm{x} - \bm{s})
\end{equation} 
for $\bm{x} \in \mathbb{Z}^2$ and $t = 0, 1, 2, \dots$.
This provides a quantum analog of the difference equation
that governs the probability distribution of a 2D random walk. 
In \eqref{state_evol}, 
the two-by-two matrices $Q_{\bm{s}}$
are assumed to be independent of the position, 
which implies that
the evolution described by \eqref{state_evol} is homogeneous
or equivalently \emph{translation invariant}.
See FIG.~\ref{fig:2D2SQW} for the motion of the 2D2SQW in $\mathbb{Z}^2$.

Ohno \cite{ohnoParameterizationTranslationInvariantTwoDimensional2018} 
provided a criterion for the translation-invariant 2D2SQW of the form \eqref{state_evol} to be unitary. 
Based on this criterion,
one can rewrite all such unitary evolutions as
\begin{equation}
    \Psi_{t + 1} = U \Psi_t, \quad t = 0, 1, 2, \dots, \label{psitut}
\end{equation} 
where $\Psi_0$ is the initial state and
the unitary evolution operator $U$ is given by
the product of conditional shift operators $S_j$
and coin operators $C_j$ ($j = 1, 2$): 
\begin{equation}
	U := S_2C_2S_1C_1. \label{defeq:ACQW}
\end{equation}
The action of $S_j$ ($j = 1, 2$)
is defined on a state $\Psi \in \mathcal{H}$ by
$(S_j\Psi)^{(\pm)}(\bm{x}) = \Psi^{(\pm)}(\bm{x} \pm \bm{e}_j)$
with the standard basis $\{\bm{e}_j\}_{j = 1, 2}$. 
The coin operators $C_j$ are taken as arbitrary
two-by-two unitary matrices,
and they are naturally identified with the operator
$I \otimes C_j$ ($j = 1, 2$) on the full Hilbert space $\mathcal{H}$
by omitting the identity $I$.
The relation $\sum_{\bm{s} \in \{(s_1, s_2) \mid s_j = -1, +1\}} Q_{\bm{s}} = C_2C_1$ between the transition matrices $Q_{\bm{s}}$ and the coin matrices $C_j$,
along with the unitarity of $C_j$, guarantees
the unitarity of the evolution $U$ and
thus the state evolution \eqref{state_evol}. 
This is the quantum analogy of probability conservation
in the classical random walk;
that is, the sum of the probability of moving from $\bm{x} - \bm{s}$ to $\bm{x}$
over all allowed displacements $\bm{s} \in \{(s_1, s_2) \mid s_j = -1, +1\}$ equals one.   

\subsection{Parameters}
In general, any two-by-two unitary matrices $C \in U(2)$ are parameterized
by one real number $r \in [0,1]$ 
and three angles $\alpha,\beta, \delta \in [-\pi, \pi)$ as
\[
    C = C(r;\alpha,\beta, \delta) 
    := e^{i \delta/2} 
    \begin{pmatrix} e^{i \alpha } r & e^{i \beta} s \\
    - e^{-i \beta} s & e^{-i \alpha} r \end{pmatrix} 
\]
with $s = \sqrt{1 - r^2}$ and $\det C = e^{i \delta}$.
The following lemma then reduces the parameters
and leads to a canonical form of the coin operator $C_j$.
The detailed proof is given in Appendix \ref{appen:pr:lem:equiv}.
\begin{lem}
\label{lem:equiv}
Fix $a_j \in [0,1]$ and $\alpha_j, \beta_j, \delta_j \in [-\pi,\pi)$ ($j=1,2$).
Let $U$ be the operator defined by \eqref{defeq:ACQW}
with coin operators given by
\[ C_j =  C(a_j; 0, 0, 0) \quad (j=1, 2), \]
and $U^\prime:= S_2C_2^\prime S_1C_1^\prime$
with 
\[ C_j^\prime =  C(a_j; \alpha_j, \beta_j, \delta_j)  \quad (j = 1, 2). \]
Then there exists a unitary operator $W$ on $\mathcal{H}$ 
such that the following hold:
\begin{itemize}
    \item[(i)] There exists a constant $\gamma \in [-\pi,\pi)$ such that
    \[ W^{\dagger} U^\prime W = e^{i\gamma} U, \]
    i.e., $U^\prime$ and $U$ are unitarily equivalent up to a constant factor; 
    \item[(ii)] Let $\Psi^\prime_t = (U^\prime)^t\Psi^\prime_0$ with $\|\Psi^\prime_0\|=1$
    and define the position $\bm{X}^\prime_t$ in the same manner as \eqref{distXt}
    with $\Psi_t$ replaced by $\Psi^\prime_t$.
    Then $\bm{X}^\prime_t$ has the same distribution as
    $\bm{X}_t$ defined by \eqref{distXt} with $\Psi_0=W^{\dagger} \Psi^\prime_0$,
    i.e.,
    \[ P(\bm{X}^\prime_t = \bm{x}) = P(\bm{X}_t = \bm{x}), \quad \bm{x} \in \mathbb{Z}^2 \]
    for any $t = 0, 1, 2,\dots$. 
\end{itemize}
\end{lem}

From Lemma \ref{lem:equiv},
the probability distribution \eqref{distXt}
of the position $\bm{X}_t$ is independent of 
the phase angles $\alpha_j, \beta_j, \delta_j \in [-\pi,\pi)$ ($j = 1, 2$).
Consequently, the limit distribution is also independent of these angles. 
Without loss of generality, 
we can, therefore, suppose that the coin operators are of
the simpler canonical form:
\begin{equation}
	C_j = C(a_j; 0, 0, 0)
    = \begin{pmatrix}
		a_j & b_j \\
		- b_j  & a_j 
	\end{pmatrix}
    \label{defeq:coin_operators}
\end{equation}
with $a_j \in [0, 1]$ and $b_j = \sqrt{1 - a_j^2}$. 
Thus, the translation-invariant 2D2SQW model treated in this paper
is fully parameterized by only two real numbers $a_1, a_2 \in [0,1]$,
which coincides with Ohno's result~\cite{ohnoParameterizationTranslationInvariantTwoDimensional2018}.  

\begin{figure}
	\includegraphics[width=4.5cm]{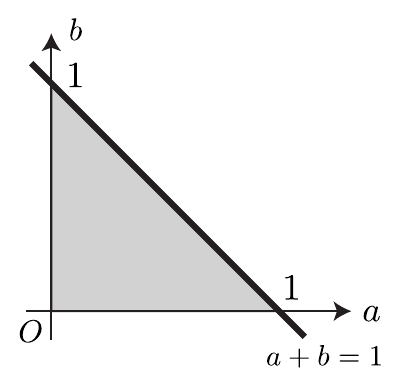}
	\caption{Parameter region for $(a, b) \in \Delta$.
    The parameter pair can take any value within the interior of the gray-shaded triangle 
    and its boundary where $a + b = 1$.
    } \label{fig:ab}
\end{figure}

To facilitate the presentation of the main results,
we parameterize our model using two quantities, 
$a$ and $b$, derived from \eqref{defeq:coin_operators}
by:
\begin{equation}
\label{def:ab}
	a := a_2 a_1,\quad  b := b_2 b_1.
\end{equation}
By definition,
the resulting pair $(a, b)$ of parameters
must be in the triangular region $\Delta$ defined by
the three constraints (depicted in FIG. \ref{fig:ab}):
\begin{equation*} 
\Delta = \{(a, b) \mid a \geq 0, \ b \geq 0,  \ a + b \leq 1\}.
\end{equation*}
The following lemma ensures that $\Delta$
provides a non-redundant parameterization of our model by establishing a one-to-one correspondence between the ordered coin parameters
$(a_1, a_2)$ and $(a, b) \in \Delta$.
The proof is deferred to Appendix \ref{appen:pr:lem:a1a2toab}.
\begin{lem} \label{lem:a1a2toab}
For $(i, j) =(1, 2)$ or $(2, 1)$,
let $A_{ij} = \{ (a_1,a_2) \mid 0 \leq a_i \leq a_j \leq 1 \}$. 
Define maps 
\[ \Phi_{ij}: A_{ij}\ni (a_1, a_2) \longmapsto (a, b) \in \Delta \] 
so that $(a, b)$ satisfies \eqref{def:ab}.
Then $\Phi_{ij}$ are one-to-one onto correspondences.     
\end{lem}

\section{Results} \label{sec:main_theorem}
This paper derives an explicit expression for the limiting PDF
of the scaled position $\bm{X}_t / t$ for translation-invariant 2D2SQWs parameterized by $(a, b)$ in the region depicted in FIG.~\ref{fig:ab}.
By Lemmas \ref{lem:equiv},
one can assume, without loss of generality,
that the evolution operator $U = S_2C_2S_1C_1$
has coin operators $C_j$ of the canonical form
\eqref{defeq:coin_operators}.
The state $\Psi_t$ of the walker at time $t = 0, 1, 2, \dots$
is given by \eqref{psitut}:
$\Psi_t = U^t \Psi_0$,
where the initial state $\Psi_0 \in \mathcal{H}$ 
is an arbitrary normalized vector.
The probability distribution $P(\bm{X}_t = \bm{x})$ is obtained from
the state $\Psi_t(\bm{x})$ via \eqref{distXt}.
Lemma \ref{lem:a1a2toab} ensures that $(a, b) \in \Delta$
fully determines the model through $a = a_1a_2$ and $b = b_1b_2$.

According to \cite{grimmettWeakLimitsQuantum2004},
the limit distribution is given by
\begin{equation*}
    P(V_1 \leq u_1, V_2 \leq u_2) =
    \int_{-\infty}^{u_1} \int_{-\infty}^{u_2} 
    d \braket <\Psi_0 | E_{\hat{\bm{v}}} ({\bm v}) \Psi_0>,
\end{equation*}
where $E_{\hat{\bm{v}}} = E_{\hat{v}_1} \otimes E_{\hat{v}_2}$ denotes 
the (joint) spectral measure
for the (pair of asymptotic) velocity operators $\hat{\bm{v}} = (\hat{v}_1, \hat{v}_2)$.
and one may formally represent
$\braket <\Psi_0 | E_{\hat{\bm{v}}}({\bm v}) \Psi_0 > $
as $|\braket < {\bm v} |\Psi_0 >|^2$
(see Sec.~\ref{sec:proofs} for the precise definition). 
Intuitively, in the wave vector space representation,
the velocity operators $\bm{v}(\bm{k}) = \bm{(}v_1(\bm{k}), v_2(\bm{k})\bm{)}$
are equal to the group velocity,
 $\bm{v}(\bm{k}) = \nabla \omega$, 
whose ranges determine the spectra $\sigma(\hat{\bm{v}}_j)$.
Here, $\omega(\bm{k})$ denotes the dispersion relation of our model,
which satisfies $\cos \omega(\bm{k}) = \tau(\bm{k})$ with 
\begin{equation} \label{defeq:tau}
    \tau(\bm{k}) = a \cos (k_2 + k_1) - b \cos (k_2 - k_1)
\end{equation}
for the wave vector $\bm{k} = (k_1, k_2) \in \mathbb{T}^2$.

\subsection{Maximal Speeds and Parameters}
As in the 1DQW model the maximal speed plays a key role in characterizing the limit distribution
and the asymptotic behavior.
We thus define the maximal speed of the 2D2SQW as
\[ v_{\mathrm{max}} := \max_{j = 1, 2} \max_{\bm{k} \in \mathbb{T}^2} |v_j(\bm{k})|. \]
Expressing $v_j$ in terms of the function $\tau(\bm{k})$ 
and its derivatives, 
we can determine the spectra $\sigma(\hat{v}_j)$
and the maximal speed $v_{\mathrm{max}}$.
The explicit result is given by the following proposition:
\begin{prop}[Maximal speed] \label{prop:hat_v_q}
    The following hold:
    \begin{itemize}
        \item[(i)] $v_{\mathrm{max}} = a + b$;
        \item[(ii)] $\sigma(\hat{v}_j) = [-v_{\mathrm{max}}, v_{\mathrm{max}}]$ for $j = 1, 2$. 
    \end{itemize}
\end{prop}

As shown later, 
the support of the limit distribution equals
the support $\Sigma(\hat{\bm{v}})$ of the joint spectral measure $E_{\hat{\bm{v}}}$. 
In general, the joint spectrum $\Sigma(\hat{\bm{v}})$ is strictly contained in the Cartesian product of the spectra $\sigma(\hat{v}_j)$
(see Sec.~\ref{subsec:jointspectrum} for more details):
$\Sigma(\hat{\bm{v}}) \subset \sigma(\hat{v}_1) \times \sigma(\hat{v}_2)$.

Di Franco et al.~\cite{difrancoAlternateTwodimensionalQuantum2011} showed that for AQWs with $v_{\mathrm{max}} = 1$, 
$\Sigma (\hat{\bm{v}})$ consists of a single ellipse $E$ given in \eqref{ellipse}.
More recently,
Cedzich et al.~\cite[Sec.~5.2]{cedzichExponentialTailEstimates2024})
reported a model, corresponding to the $a=b=1/2$ case
of our general framework,
where $\Sigma (\hat{\bm{v}})$ becomes a unit circle
and $v_{\mathrm{max}}=1$.
As suggested in \cite[FIG.~4~(b)]{ahlbrechtAsymptoticEvolutionQuantum2011}, 
for an AQW with $v_{\mathrm{max}}<1$,
$\Sigma(\hat{\bm{v}})$ appears to be the intersection of two elliptical regions. 
These facts suggest the necessity of dividing the parameter region of $(a,b)$ 
into several distinct cases.
We thus define three regions $\Delta_{ab = 0}$,  $\Delta_{(0, 1)}$,
and $\Delta_1$ of the parameters $\Delta$ as 
\begin{align*}
    & \Delta_{ab = 0} = \{ (a, b) \in \Delta \mid ab = 0 \}, \\
    & \Delta_{(0, 1)} = \{ (a, b) \in \Delta \mid \ a + b < 1, \ ab \neq 0 \},  \\
    & \Delta_1 = \{ (a, b)\in \Delta \mid a + b =1,
    \ ab \neq 0 \}. 
\end{align*}
\subsection{Regime of Reducible QWs: \texorpdfstring{$\Delta_{ab = 0}$}{Delta {ab = 0}}}
\label{sec:reducibleQWs}
If the parameters satisfy $(a, b) \in \Delta_{ab = 0}$,
our model can be reduced to two simpler scenarios:

(i) The walker exhibits $v_{\mathrm{max}} = 0$ if and only if
$a = b = 0$. 
In this case, the walker is bound in a finite region,
which corresponds to (a) of TABLE \ref{tab:1} for the 1D2SQW.

(ii) The walker has nonzero velocity, i.e.,
$v_{\mathrm{max}} \in (0,1)$,
but the parameters are supposed
to satisfy either $a=0$ or $b=0$.
For instance, in the case where $a_1=0$, i.e., $a=0$ and $b=b_2$,
one can prove, by a method similar
to Lemma \ref{lem:equiv}, that
$U$ is unitarily equivalent to the decoupled form:
\[
    U \simeq\bigoplus_{n \in \mathbb{Z}} S_{\mathrm{1D}}\tilde C_2
    \quad \text{on}
    \quad \bigoplus_{n \in \mathbb{Z}} \ell^2(\mathbb{L}_n^{(-)}; \mathbb{C}^2),
\]
where the walker's motion is constrained to
each line $\mathbb{L}_n^{(-)}$ ($n \in \mathbb{Z}$).
Here we use $\mathbb{L}_n^{(\pm)}$
to denote the line:
\[ 
    \mathbb{L}_n^{(\pm)} =\{(m, n \pm m) \mid m \in \mathbb{Z} \} 
    \subset \mathbb{Z}^2
    \quad (n \in \mathbb{Z})
\]
and identify $\mathbb{L}_n^{(\pm)}$
with $\mathbb{Z}$ in a natural way such that $(m, n\pm m) \mapsto m$.
In the above decomposition,
the action of the operator $S_{\mathrm{1D}} = S_2^{\dagger}S_1$
on $\ell^2(\mathbb{L}_n^{(-)}; \mathbb{C}^2)$
is represented as 
$(S_{\mathrm{1D}}\Psi)^{(\pm)}(m)=\Psi^{(\pm)}(m \pm 1)$. Also, the symbol $\tilde{C}_j$ denotes $C(b_j; \pi, 0, 0)$. 

TABLE \ref{tab:3} summarizes the complete set of
decompositions for all choices in $\Delta_{ab = 0}$,
confirming that the resulting 1D distributions
on each line are given established 1D WLT results
in TABLE \ref{tab:1}.
\begin{table}[ht]
    \centering
    \begin{tabular}{|l||c|c|c|}
        \hline
        Choice & $v_{\mathrm{max}}$ & 1D decomposition & Lines \\ \hline
        $a_1 = 1$ 
            & $a = a_2$ 
            & $\tilde{S}_{\mathrm{1D}}C_2$ 
            & $\mathbb{L}_n^{(+)}$ \\ \hline
        $a_2=1$ 
            & $a=a_1$ 
            & $\tilde{S}_{\mathrm{1D}}C_1$ 
            & $\mathbb{L}_n^{(+)}$ \\ \hline
        $a_1=0$ 
            & $b=b_2$ 
            & $S_{\mathrm{1D}} \tilde{C}_2$ 
            & $\mathbb{L}_n^{(-)}$ \\ \hline
        $a_2=0$ 
            & $b=b_1$ 
            & $S_{\mathrm{1D}}^{\dagger} \tilde{C}_1 $
            & $\mathbb{L}_n^{(-)}$ \\ \hline
    \end{tabular}
    \caption{
        Complete list of choices of $(a_1, a_2)$ 
        corresponding to $(a, b)\in \Delta_{ab=0}$.
        The action of 
        $\tilde{S}_{\mathrm{1D}}=S_2S_1$ on  $\ell_2(\mathbb{L}_n^{(+)};\mathbb{C}^2)$ 
        parallels the action of $S_{\mathrm{1D}}$ on 
        $\ell_2(\mathbb{L}_n^{(-)};\mathbb{C}^2)$. 
    }
    \label{tab:3}
\end{table}

In subsequent sections, 
we therefore focus our analysis
exclusively on the case where $ab \ne 0$,
which corresponds to the intrinsically 
2D motion of the walker.

\subsection{\texorpdfstring{
Regime of Konno Functions:
$\Delta_{(0, 1)}$
}{Regime of Konno Functions: Delta (0, 1)}} 
\label{subsec:mainA}
This subsection supposes that:
\[
    (a,b) \in \Delta_{(0,1)},
    \quad \text{i.e.,} \quad
    v_{\mathrm{max}} \in (0,1), \quad ab\not=0.
\]
Let $F_j$ ($j=1,2$) and $F_0$ be functions given by
\begin{align}
    F_j({\bm v}) 
    & = 1 - \dfrac{(v_1 + v_2)^2}{4a_j^2} - \dfrac{(v_1 - v_2)^2}{4b_j^2}, \notag \\
    F_0({\bm v})
    & = \dfrac{(a_1b_2)^2+(a_2b_1)^2}{2ab} \notag \\
    & \times \left[ 1 -\dfrac{(v_1 + v_2)^2}{4a_0^2} 
     - \dfrac{(v_1-v_2)^2}{4 b_0^2}
     \right]
     \label{eq:B_with_v} 
\end{align}
with
\[ 
a_0^2 
= \frac{(a_1b_2)^2 + (a_2b_1)^2}{b_1^2+b_2^2}, 
\quad b_0^2 
= \frac{(a_1b_2)^2 + (a_2b_1)^2}{a_1^2 + a_2^2}
\]
and denote by $E_j$ the ellipses defined by $F_j$:
\begin{equation} \label{ellipses}
	E_j = \left\{ {\bm v} =(v_1, v_2) \,\middle|\, 
    F_j({\bm v}) \geq 0
    \right\}.
\end{equation}

We now present our main result:
\begin{thm}[2D Konno Functions] \label{thm:WLT}
Suppose that $(a, b) \in \Delta_{(0, 1)}$. 
Then the limiting distribution of $\bm{X}_t / t$ is given by
\begin{align*}
    & P(V_1 \leq u_1, V_2 \leq u_2)  \notag \\
    & \quad \qquad = \int_{-\infty}^{u_1}  \int_{-\infty}^{u_2} dv_1 dv_2 
    \sum_{\bullet = \pm} w_{\bullet} ({\bm v})  f_{\bullet} ({\bm v}),
\end{align*} 
where $w_{\pm}$ are weight functions defined in \eqref{eq:w_bullet},
and $f_{\pm}$ are state-independent functions defined as
\begin{equation}
    f_\pm = \dfrac{F_0 \pm \sqrt{F_1 F_2}}{2\pi^2 (1 - v_1^2)(1 - v_2^2)\sqrt{F_1 F_2}}
    I_{E_1 \cap E_2}. 
    \label{eq:fpm}
\end{equation}
\end{thm}

\begin{figure}[tb]
    \centering
    \includegraphics[width=7cm]{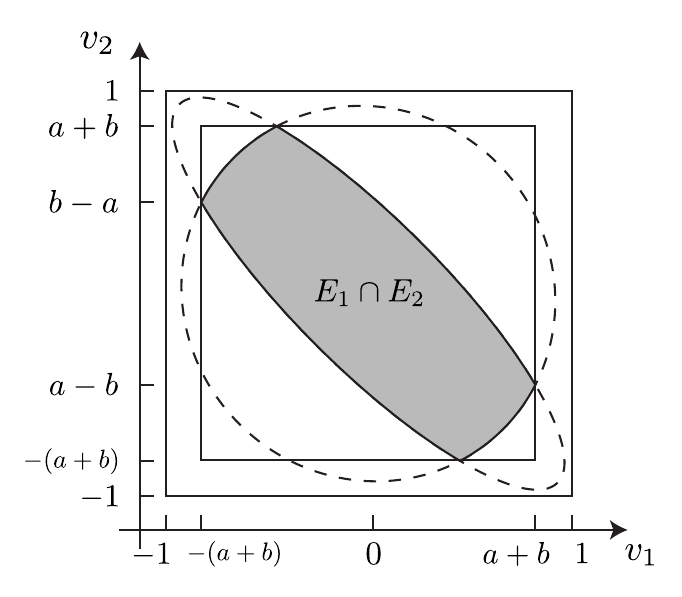}
    \caption{The support of 2D Konno functions $f_{\pm}$ in \eqref{eq:fpm}
    is $E_1\cap E_2,$ which is the intersection of two ellipses (whose boundaries are shown as dashed lines) and is colored gray. The region is plotted for $(a_1,a_2)=(1/4, 3/4),$ i.e., $(a, b) = (3/16, \sqrt{105}/16).$ In general, the domain $\Sigma(\hat{\bm{v}}) = E_1\cap E_2$ is included in $\sigma(\hat{v}_1)\times \sigma(\hat{v}_2) = [-(a + b), a + b]^2$, which is also confirmed in this figure. In particular, $E_1 \cap E_2$ is contained in $[-1, 1]^2$, as visually confirmed in this figure, and therefore Lemma ~\ref{lem:5}(3) holds.} \label{fig:E1_cap_E2}
\end{figure}

\begin{figure*}[bt]
    \centering
    \includegraphics[width=14cm]{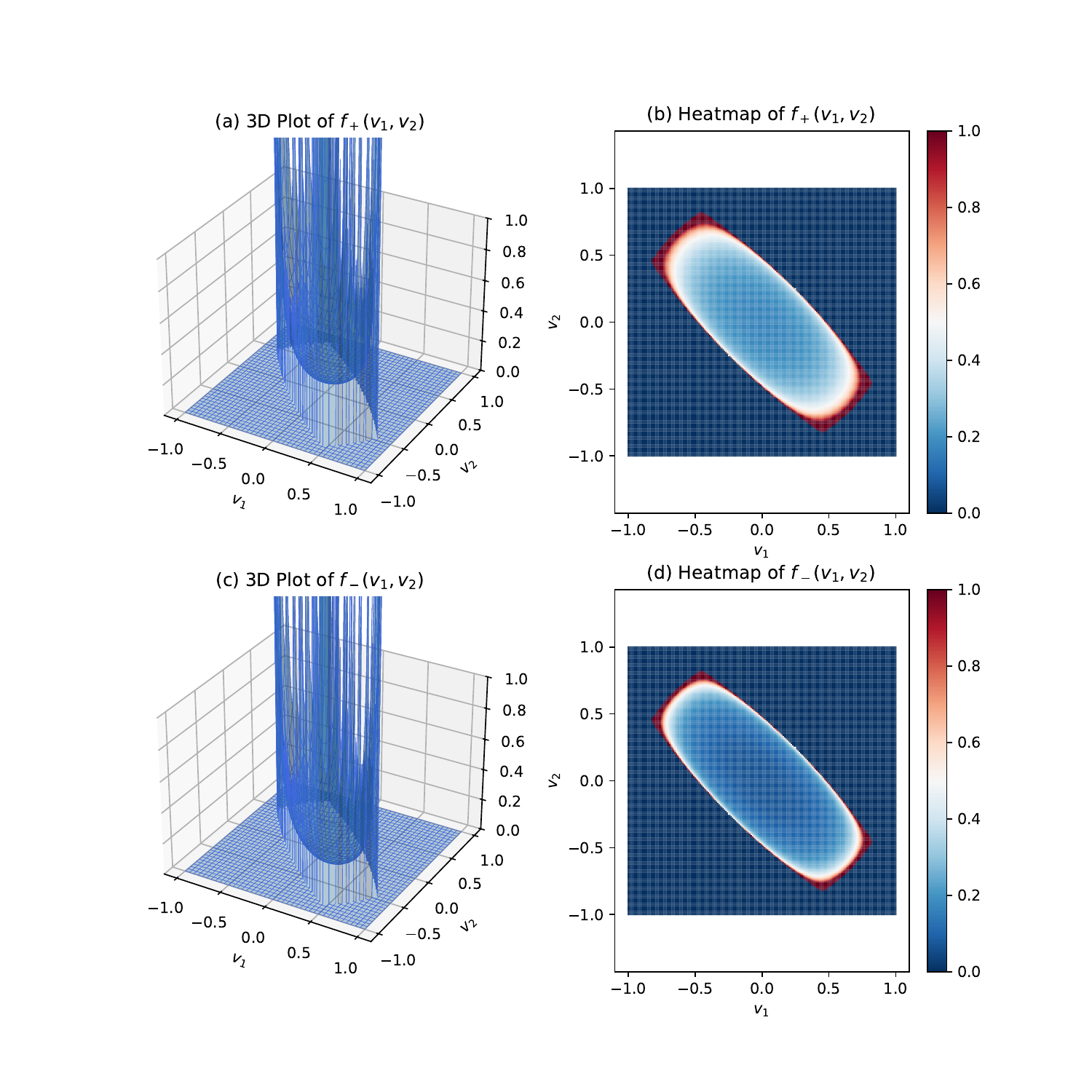}
    \caption{(a) and (c) show 3D plots of 2D Konno functions $f_+({\bm{v}})$ and $f_-({\bm{v}})$, respectively, for $(a_1, a_2)=(1/4, 3/4),$ i.e., $(a, b) = (3/16, \sqrt{105}/16)$, similar to FIG.~\ref{fig:E1_cap_E2}. (b) and (d) show the corresponding heat maps. In the heat maps, the color scale represents the function's value, which is zero (dark blue) outside the domain and increases toward the boundary (light blue to white/red), as indicated by the color bars. The set in (b) and (d) is identical to that in FIG.~\ref{fig:E1_cap_E2}, and it can be seen that $f_{\pm}({\bm{v}})$ diverges on this boundary. 
    This fact corresponds to the divergence of 1D Konno function $f_{\mathrm{K}}$ at the points $v = \pm v_{\mathrm{max}}$.
    } \label{fig:plot_fpm}
\end{figure*}

We sketch the proof of Theorem~\ref{thm:WLT} in Sec.~\ref{sec:proofs}.
From Lemma~\ref{lem:5} (proved in Appendix~\ref{app:pr:lem:5}) below,
$f_{\pm}$ are well defined, strictly positive in the interior of $E_1 \cap E_2$,
and $f_{\pm}$ diverge only on the boundary of $E_1 \cap E_2$.

\begin{lem} \label{lem:5}
Let $(a, b) \in \Delta_{(0, 1)}$. 
\begin{itemize}
    \item[(1)] $F_1F_2 \geq 0$ in $E_1 \cap E_2$
        and $F_1F_2 = 0$ only on the boundary of $E_1 \cap E_2$;
    \item[(2)] $F_0 > \sqrt{F_1F_2}$ in $E_1 \cap E_2$;
    \item[(3)] $(1 - v_1^2)(1 - v_2^2) > 0$ in $E_1 \cap E_2$.
\end{itemize}
\end{lem}

FIG.~\ref{fig:E1_cap_E2} depicts the domain of $f_{\pm}$. In particular, the domain of the PDF is $E_1 \cap E_2$, independent of the initial state. FIG.~\ref{fig:plot_fpm} shows the 3D plot and the heat map of $f_{\pm}$. Especially, the fact that $f_{\pm}$ diverges on the boundary of $E_1 \cap E_2$ is consistent with Lemma~\ref{lem:5}.

\subsubsection*{1D limit to Konno function}
To establish the connection between our 2D functions $f_{\pm}$ and the celebrated 1D Konno function \eqref{KonF1}, we define the appropriate reduction
of our 2D model in the 1D limit. 
To make the dependence on $(a, b) \in \Delta$ clear,
we use $U_{(a, b)}$ to denote the evolution operator $U = S_2C_2S_1C_1$ and
we also denote the limiting PDF obtained in Theorem \eqref{thm:WLT}
by
$\rho_{(a,b)}({\bm v})
    = \sum_{\bullet = \pm} w_{\bullet} ({\bm v})  f_{\bullet} ({\bm v})$.
We use the notation from Lemma \ref{lem:a1a2toab}. 
\begin{dfn}[1D Limit]
\label{def:1DL(01)}
Take $(a_1, a_2) = \Phi_{12}^{-1}\bm{(}(a, b)\bm{)}$ so that $(a, b) \in \Delta_{(0, 1)}$ and fix $a_1$. 
We say that $\rho_{(a, b)}$ has the 1D limit if 
there exists a 1D PDF $\rho_{a_1}(v)$ such that
\begin{equation}
\label{defeq:1D(01)}
\lim_{b_2 \to 0} \int_{\Sigma(\hat{\bm v})} 
    g({\bm v})\rho_{(a,b)}({\bm v}) d{\bm v}
    = \int_{-a_1}^{a_1} g(v,v) \rho_{a_1}(v) dv
\end{equation}
for any bounded continuous function $g({\bm v})$.
In this case, we write the limit as
$\lim_{b \to 0} \rho_{(a,b)} = \rho_{a_1}$
and call it
the 1D limit of $\rho_{(a, b)}$ as $b \to 0$.
\end{dfn}
The 1D limit procedure
involves the following:
\begin{itemize}
    \item[(1)] The parameters $(a, b)$ remain strictly
    within the regime $\Delta_{(0,1)}$ as they approach the boundary $\Delta_{ab = 0}$,
    corresponding to $b_2 \to 0$ from above.
    \item[(2)] Due to the limit $\lim_{b \to 0}(a, b)=(a_1, 0)$, the maximal speed $v_{\mathrm{max}} = a + b$ has a nonzero limit $a_1 \in (0, 1)$.
    \item[(3)] The evolution $U_{(a, b)}$ converges to
    $U_{a_1}:=S_2S_1C_1$, which represents a 1D motion along each line $\mathbb{L}_n^{(+)}$ ($n \in \mathbb{Z}$) 
    (see TABLE \ref{tab:3}).
    \item[(4)] The RHS of \eqref{defeq:1D(01)} can formally be written as
    \begin{equation} 
    \label{distkernel}
    \int_{\Sigma(\hat{\bm v})} g({\bm v})\delta(v_2 - v_1) \rho_{a_1}\left( v_1 \right)\, d{\bm v}, 
    \end{equation}
    which implies that
    the 2D PDF $\rho_{(a,b)}$ degenerates
    in the 1D limit $b \to 0$.
\end{itemize}

The following theorem 
(proved in Appendix.~\ref{sec:1Dlimits})
implies that our function $f_\pm$
converges to the 1D Konno function $f_{\mathrm{K}}$ in the 1D limit.
\begin{thm}[1D Limit]\label{thm:btozero}
Let $(a, b) \in \Delta_{(0, 1)}$. Then
$\rho_{a_1}:= \lim_{b \to 0}\rho_{(a, b)}$ exists and
\begin{equation} \label{rhsG_0}
    \rho_{a_1}(v)
    = w(v; a_1) f_{\mathrm{K}}(v; a_1),
\end{equation}
where the weight function $w(\cdot;a_1)$ is defined
in \eqref{eq:weight_bto0}. 
\end{thm}
This result confirms that our derived functions
actually constitute the 2D generalization of the
1D Konno distribution, and therefore we refer to
$f_\pm$ as the 2D Konno functions.
We discuss how the 2D functions $f_\pm$ reduce
to the Konno function $f_{\mathrm{K}}$ 
at the end of Sec.~\ref{subsec:interpretation}. 

\subsection{\texorpdfstring{Regime of Maximal Speed $v_{\mathrm{max}} = 1$: $\Delta_1$}{Regime of Maximal Speed vmax = 1: Delta 1}}

This subsection assumes that:
\[ 
    (a,b) \in \Delta_1, \quad
    \text{i.e.}, \quad v_{\mathrm{max}} = 1, \quad ab \neq 0.
\]
Eqs.~\eqref{def:theta1} and \eqref{def:theta2} in Appendix \ref{appen:pr:lem:a1a2toab} prove that
$a + b = 1$ if and only if $a_1 = a_2$ and $b_1 = b_2$.
Recall that $F_j = 0$
defines the boundary of the ellipse $E_j$  ($j = 1, 2$)
by \eqref{ellipses}.
Similarly, $F_0$ defined in \eqref{eq:B_with_v}
also defines a boundary of an ellipse.
The function
\[
    F({\bm v}) 
    = 1 - \frac{(v_1 + v_2)^2}{4a} - \frac{(v_1 - v_2)^2}{4b}
\]
gives the boundary of the ellipse $E$
defined in \eqref{ellipse}.
\begin{lem} \label{lem:EFaresame}
Let $(a, b) \in \Delta_1$. Then:
\begin{itemize}
    \item[(1)] $F = F_j$ and hence $E = E_j$ ($j = 1, 2$).
    \item[(2)] $F = F_0$.
\end{itemize}
\end{lem}
\begin{proof}
(1) is obvious
from the relations $a = a_1^2 = a_2^2$ and $b = b_1^2 = b_2^2$.
These relations also provide
    $a = a_0^2$, $b = b_0^2$,  
    and $2ab = (a_1b_2)^2 + (a_2b_1)^2$.
Substituting these relations into \eqref{eq:B_with_v}
yields (2).
\end{proof}
Combining Lemma \ref{lem:EFaresame} with \eqref{eq:fpm}, 
we observe that $f_+$ coincides with $f$ defined in \eqref{KonF2},
while $f_-$ vanishes:
\begin{align*}
    f_+ = f, \quad f_-=0.
\end{align*}
Lemma \ref{lem:EFaresame} also implies that 
the intersection of two ellipses $E_1 \cap E_2$,
which constitutes the support of the 2D Konno function
$f_+$, degenerates into the single ellipse $E$ defined in \eqref{ellipse}: $E_1 \cap E_2 =  E$.

Recall that $f$ is the function defined in \eqref{KonF2}.
\begin{thm} \label{thm:WLT'}
Let $(a, b) \in \Delta_1$.
Then the limit distribution of $\bm{X}_t / t$ is given by 
\begin{align*}
    & P(V_1 \le u_1, V_2 \le u_2)  \notag \\
    & \qquad \qquad =
    \int_{-\infty}^{u_1}  \int_{-\infty}^{u_2} dv_1 dv_2 \,
    w_+ ({\bm v})  f ({\bm v}),
\end{align*} 
where the weight function $w_{+}$ is defined in Eqs.~\eqref{eq:w_bullet}. 
\end{thm}
Thus, we have recovered the function $f$
obtained
by Watabe et al.~\cite{watabeLimitDistributionsTwodimensional2008},
Di Franco et al.~\cite{difrancoAlternateTwodimensionalQuantum2011}
as a special case of $f_+$
with $v_{\mathrm{max}}=1$ and $ab \neq 0$. 
Although the statement of Theorems \ref{thm:WLT'}
might appear to be a straightforward corollary of Theorems \ref{thm:WLT} 
in light of Lemma \ref{lem:EFaresame}, 
the proof requires
a careful approach
due to the discontinuity of the group velocity $\bm{v}(\bm{k})$.
This kind of difficulty was previously noted in \cite[Technical remark at the end of Sec.~III]{grimmettWeakLimitsQuantum2004}. 
We address this problem 
at the end of Sec.~\ref{sec:DerivationPDF}.

\subsubsection*{Triviality of the 1D limit}
While the function $f_+=f$ in the $\Delta_1$ regime recovers the prior studies
\cite{watabeLimitDistributionsTwodimensional2008,difrancoAlternateTwodimensionalQuantum2011},
the following theorem reveals that
$f$ \emph{cannot} be properly interpreted as a genuine 2D generalization of the Konno function; 
the 1D limit of the PDF in the $\Delta_1$ regime becomes \emph{trivial}. 

Note that for $(a, b) \in \Delta_1$,
the relation $b_2 = b_1 = \sqrt{b}$ holds,
unlike the case for $(a, b) \in \Delta_{(0, 1)}$.
Hence, $(a, b)$ converges to $(1, 0)$
as $b \to 0$ (or equivalently, $b_1 = b_2 \to 0$ and $a_1=a_2 \to 1$). 
With this procedure, the evolution
$U_{(a, b)} = S_2C_2S_1C_1$
converges to $U_1 = S_2S_1$,
which is the 1D2SQW moving only in a single direction.

From these observations, Definition~\ref{def:1DL(01)} requires the following modification for $(a,b) \in \Delta_1$:
Let $\rho^{(1)}_{(a,b)}=w_+ f$ be the PDF
obtained by Theorem~\ref{thm:WLT'}. 
We say that $\rho_{(a, b)}^{(1)}$ has the 1D limit if 
there exists a 1D distribution
(generalized function)
$\rho_{1}(v)$,
understood as a measure,
such that
\begin{equation*}
\lim_{b_2 \to 0} \int_{\Sigma(\hat{\bm v})} 
    g({\bm v})\rho_{(a,b)}^{(1)}({\bm v}) d{\bm v}
    = \int_{-1}^{1} g(v,v) \rho_{1}(v) dv
\end{equation*}
for any bounded continuous function $g({\bm v})$.
Similarly to \eqref{distkernel},
the integral on the RHS is formally interpreted as the action of the distribution
$\delta(v_2-v_1)\rho_1(v_1)$ on a test function $g$.
\begin{thm} \label{thm:btozero1}
Let $(a, b) \in \Delta_1$.
Then $\rho_1 := \lim_{b \to 0}\rho_{(a, b)}^{(1)}$ exists
and
\begin{equation}
\label{eq:1D(1)limitrepr}
\rho_1(v) =
        C_{-1} \delta_{-1}(v) + C_{+1} \delta_{+1}(v), 
\end{equation}
where the constants $C_{\pm 1} = \left\|\Psi_0^{(\mp)}\right\|_{\ell^2(\mathbb{Z}^2)}^2$.
\end{thm}
Theorem~\ref{thm:btozero1} (proved in Appendix~\ref{sec:1Dlimits}) means that the 2D PDF $w_+ f$ degenerates
to the 1D distribution corresponding to the trivial 
ballistic motion with $v_{\mathrm{max}} = 1$ (see TABLE \ref{tab:1} (b)).
We thus find that the function $f$ (recovered from prior studies
\cite{watabeLimitDistributionsTwodimensional2008,difrancoAlternateTwodimensionalQuantum2011})
is a 2D generalization of the \emph{trivial} 1D ballistic motion
($v_{\mathrm{max}} = 1$).
This contrasts sharply with our 2D Konno functions $f_\pm$, 
which we have shown to be the genuine generalization of 
the \emph{non-trivial} 1D Konno distribution ($v_{\mathrm{max}} < 1$).


\section{Discussion} \label{sec:discussion}
This paper has dealt with 2D2SQW in a general form,
successfully deriving the explicit limit PDF
of $\bm{X}_t/t$,
and clarifying its structural relation to the 1D2SQW.
TABLE \ref{tab:2} summarizes the resulting distribution
and their 1D limits for each parameter region in $\Delta$. 

\begin{table}[ht]
    \centering
    \begin{tabular}{|l||c|c|c|c|c|}
        \hline
        $(a, b)$ & $v_{\mathrm{max}}$ & Distributions & 1D Limits  \\ \hline
        $\Delta_{ab = 0}$ & $\max\{a, b\} $ & 1D2SQW & \diagbox[dir=NE]{ \quad \quad \qquad }{ } \\ \hline
        $\Delta_{(0, 1)}$ & $0 < a + b < 1$ & $w_+f_+ + w_-f_-$ & $w f_{\mathrm{K}}$  \\ \hline
        $\Delta_1$ & $1$ & $w_+ f$ & $C_{+1}\delta_{+1} + C_{-1}\delta_{-1}$  \\ \hline
    \end{tabular}
    \caption{Limit distributions and 1D limits}
    \label{tab:2}
\end{table}

\subsection{How 2D Functions Reduce to Konno Function} \label{subsec:interpretation}

We have demonstrated in the previous section
that the function $f$ converges to trivial 1D dynamics
in the 1D limit, while our 2D function $f_{\pm}$ converges
to the Konno function $f_{\mathrm{K}}$.
The key distinction between $f_\pm$ and $f$ arises from
the presence of the factor:
\begin{equation*} 
    K_\pm({\bm v}; v_{\mathrm{max}}) = \frac{F_0 \pm \sqrt{F_1F_2}}{2\sqrt{F_1F_2}} I_{E_1 \cap E_2}.
\end{equation*}
When $v_{\mathrm{max}} = 1$,
these factors formally become 
\[ K_+({\bm v}; 1) = I_E({\bm v}), \quad K_-({\bm v}; 1) = 0, \]
leading to $E_1 \cap E_2 = E$, $f_+ \equiv f$ and $f_- \equiv 0$
(see Lemma \ref{lem:EFaresame} and the surrounding discussion). 

\begin{figure}
	\includegraphics[width=7cm]{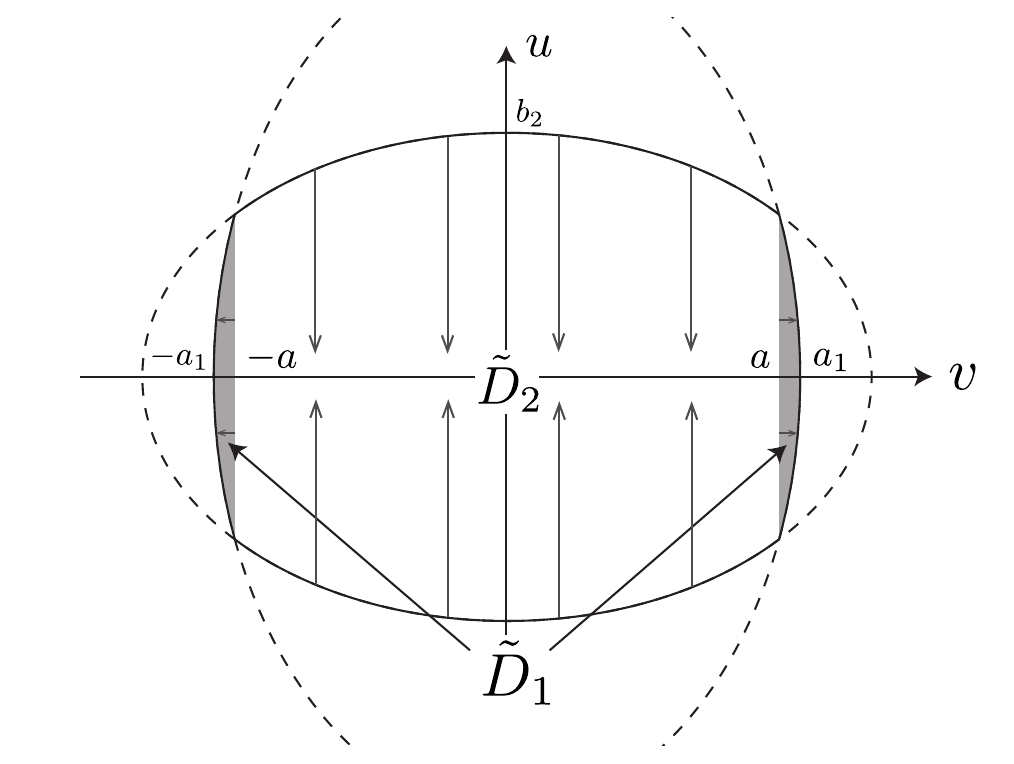}
	\caption{Domain deformation towards the 1D limit. 
        The support of the 2D Konno functions $f_\pm$
        is depicted in the rotated coordinate system $(v,u)$,
        which is decomposed into $\tilde{D}_1$ and $\tilde{D}_2$.
        The region $\tilde{D}_1$ collapses to points
        $(v, u) = (\pm a_1, 0)$, while $\tilde{D}_2$ shrinks to the 1D interval $[-a_1, a_1]$, which coincides with the support of $f_{\mathrm{K}}$.
    } \label{fig:domain_def}
\end{figure}

The factor $K_{\pm}$ also
provides a key to understanding
the mechanism behind the convergence of the 2D Konno function $f_{\pm}$ to the 1D Konno function
$f_{\mathrm{K}}$ in the 1D limit $b \to 0$.
To analyze this limit,
we introduce a change of variables that rotates the coordinate
system by $\pi / 4$:
$v = (v_1 + v_2)/2$, $u =(v_1 - v_2) / 2$.
We define
\[ \tilde{K}_\pm(v, u) = K_\pm(v + u, v - u; v_{\mathrm{max}}). \]
The boundary functions $F_j$ ($j = 0, 1, 2$) are thus
transformed into
\begin{align*} 
    \tilde{F}_j(v,u)
    := F_j(v + u, v - u) 
    = 1 - \frac{v^2}{a_j^2} - \frac{u^2}{b_j^2}.
\end{align*}
This transformation reduces the support of the functions
$f_{\pm}$ (the intersection $E_1 \cap E_2$) into
$\tilde{E}_1 \cap \tilde{E}_2$ in the $vu$-plane.
Assuming the ordering
$0 < b_2 < b_1 < 1$ and $0 < a_1 < a_2 < 1$,
we observe that
$\tilde{E}_1 \cap \tilde{E}_2$ is decomposed into
a disjoint union:
$\tilde{E}_1 \cap \tilde{E}_2 = \tilde{D}_1 \cup \tilde{D}_2$,
where $\tilde{D}_1= \left\{ (v,u) \mid a < |v| \leq a_1, \ |u| \leq r_1 \right\}$ and
$\tilde{D}_2= \left\{ (v, u) \mid |v| \leq a, \ |u| \leq r_2 \right\}$
with
$r_j = (b_j / a_j) \sqrt{a_j^2 - v^2}$ ($j = 1, 2$).
As illustrated in FIG.~\ref{fig:domain_def}, 
the fate of these two regions as the 1D limit $b \to 0$
determines the convergence to the Konno function $f_{\mathrm{K}}$.

This argument is justified by the following proposition,
which is proved in Appendix \ref{app:pr:prop:f_Kmetges}.
\begin{prop} \label{prop:f_Kemerges}
Let $(a, b) \in \Delta_{(0, 1)}$.
For any bounded continuous function $\tilde{h}$,
the following holds:
\begin{align*} 
    & \lim_{b \to 0} \iint_{\tilde{D}_1} \tilde{h} \tilde{K}_\pm \, dvdu = 0, \\
    & \lim_{b \to 0} \iint_{\tilde{D}_2} \tilde{h} \tilde{K}_\pm \, dvdu
    = \dfrac{\pi b_1}{4}   \int_{-a_1}^{\ a_1} 
    \dfrac{1 - v^2}{\sqrt{a_1^2 - v^2}} \tilde{h}(v, 0) \, dv.
\end{align*}
\end{prop}

We examine the implications of Proposition \ref{prop:f_Kemerges}.
Indeed, rotating coordinates by $\pi / 4$ gives
\begin{equation*}
    f_{\pm}(v + u, v - u) 
    = \dfrac{\tilde{K}_{\pm}(v, u)}{\pi^2\bm{(}1 - (v + u)^2\bm{)}\bm{(}1 - (v - u)^2\bm{)} }.
\end{equation*}
Combining this with Proposition \ref{prop:f_Kemerges}
provides that $f_{\pm} \, dv_1dv_2$ converges to
\begin{align*}
    & \dfrac{\pi b_1}{4} 
    \left(\dfrac{1 - v^2}{\sqrt{a_1^2 - v^2}} I_{[-a_1, a_1]}(v) \right)
    \times \frac{2\delta_0(u)\, dv du}{\pi^2(1 - v^2)^2} \notag \\
    & = f_{\mathrm{K}}(v; a_1) \times \frac{\delta_0(u)\,dvdu}{2} 
\end{align*} 
in the sense that
\begin{equation*}
    \lim_{b \to 0} 
    \iint h f_\pm\, dv_1dv_2 
    = \frac{1}{2}  \int h(v_1, v_1) f_{\mathrm{K}}(v_1; a_1)\, dv_1
\end{equation*} 
for any bounded continuous function $h(v_1,v_2)$.

To the best of our knowledge, this is the first instance in which the 1D Konno function $f_{\mathrm{K}}$ has been derived from a higher-dimensional model by taking a suitable limit, solidifying the role of $f_{\pm}$ as the genuine 2D Konno functions.
Note that this argument demonstrates the convergence mechanism for the state-independent part $f_\pm$; the rigorous proof of Theorem~\ref{thm:btozero} for the entire PDF $\rho_{(a,b)}$, which requires the precise definition of the weight functions $w_\pm$ in Sec. V, is provided in Appendix~\ref{sec:1Dlimits}.

\subsection{Analytical resolution of singular asymptotics}
\label{subsec:singularasymptotics}

The explicit derivation of the 2D Konno functions $f_{\pm}$ offers 
the first analytical understanding of the singular asymptotics of 2DQWs, a phenomenon previously addressed mainly in geometrical and phenomenological terms
\cite{baryshnikovTwodimensionalQuantumRandom2011,
ahlbrechtAsymptoticEvolutionQuantum2011}.

While Baryshnikov et al.~\cite{baryshnikovTwodimensionalQuantumRandom2011}
characterized the support of limiting distributions
of 2DQWs in terms of the \emph{Gaussian curvature} of an algebraic variety, 
neither the explicit boundary of the PDF's support nor the full expression for the PDF had been obtained.

In general, the \emph{inverse effective mass tensor}
is defined by 
the Hessian matrix $\nabla \otimes \nabla \omega$ of the dispersion relation $\omega$,
which is, therefore, equal to 
the Jacobian matrix $\nabla \bm{v}$ of the group velocity
$\bm{v} = \nabla \omega$. 
Consequently, the singularities arise from the infinite effective mass that causes the accumulation of wave packets in the band structure.
Ahlbrecht et al.~\cite{ahlbrechtAsymptoticEvolutionQuantum2011}
demonstrated that for a special 2D2SQW model, 
the loci, referred to as \emph{caustics}, 
where the Jacobian $J(\bm{k}) = \det (\nabla \bm{v})$ of the group velocity vanishes, presents the boundary of the PDF's domain.

Increasing dimensionality drastically increases the mathematical
and physical challenge of non-injectivity.
In the 1D case, as seen in the Introduction,
each velocity $v_{\pm}$ corresponding to frequencies $\pm \omega$
requires splitting the wave number space into two domains
where $v_\pm$ are injective and have their inverses.
The effective mass $m_{\mathrm{eff}} = (d^2 \omega / dk^2)_{k = 0}^{-1}$,
which appeared in \eqref{1Dw(v)startfromorigin}, is always finite
and presents no challenges in the regime $v_{\mathrm{max}} \in (0, 1)$.

In contrast, in 2D, the velocity still exhibits two modes, 
each of which, however, defines a two-variable transformation 
$\bm{(}v_{\pm, 1}(\bm{k}), v_{\pm, 2}(\bm{k})\bm{)}$
from $\mathbb{T}^2$ to $\Sigma(\hat{\bm{v}}) \subset \mathbb{R}^2$. 
As will be seen in Sec.~\ref{sec:proofs},
this transformation partitions the wave vector space
into 64 invertible regions,
resulting in 128 combinations for the two modes.
This inherent complexity underscores that the partition boundaries
are not mere technicalities, but also explicitly represent
the caustic loci themselves, where the Jacobian vanishes and the effective mass diverges. 

Our contribution, the successful construction of the inverses of these non-injective maps, is thus clarifying this universal singular structure 
of effective mass divergence.
Specifically, the caustic loci displayed as red lines in
\cite[FIG.~4(a)]{ahlbrechtAsymptoticEvolutionQuantum2011}
for a particular 2D2SQW is rigorously derived in our paper 
for a more general model,
appearing as the partitioning curves to define the invertible domain,
which are depicted in our FIG.~\ref{fig:windmill_waved_lines}
as arccosine curves on a windmill. 
While Ahlbrecht~et al.~stated that the caustics \emph{``could be calculated''}, their statement relied on numerical demonstration
for a specific 2D2SQW model,
rather than providing a universal analytical solution.
In contrast, our rigorous analysis, built on the intricate algebraic-geometrical partitioning of the wave vector space,
naturally determines the loci of caustics,
which precisely corresponds to the boundary of the PDF's domain,
or equivalently the joint spectrum $\Sigma(\hat{\bm{v}})$ of the velocity operators.
See Sec.~\ref{subsec:jointspectrum} for more details.

\subsection{Comparison with Other 2D Models} 
\label{subsec:correspondence_with_prior_research}
\subsubsection{Generalized Grover walks in 2D}
Watabe~et al.~\cite{watabeLimitDistributionsTwodimensional2008} analytically derived the explicit expression of the limit distribution for
a 2D four-state QW model, 
referred to as a \emph{generalized Grover walk} (GGW),
with a generalized four-state Grover coin
\cite{inuiLocalizationTwodimensionalQuantum2004a}:
\begin{equation} \label{Grovercoin}
    C_{\mathrm{G}}
    = \begin{pmatrix}
        q \Lambda -1 & \sqrt{pq} \Lambda \\
        \sqrt{pq} \Lambda & p\Lambda -1
    \end{pmatrix}
    \quad \text{with} \quad 
    \Lambda = \begin{pmatrix}
        1 & 1 \\ 1 & 1
    \end{pmatrix},
\end{equation}
where $p, q \geq 0$ and $p + q = 1$.
Identifying the Hilbert space $\ell^2(\mathbb{Z}^2; \mathbb{C}^4)$ of states for this model 
with $\mathcal{H} \oplus \mathcal{H}$ 
allows us to represent the shift operator $S_{\mathrm{G}}$ for this model 
as $S_{\mathrm{G}} = S_1 \oplus S_2$,
where $\mathcal{H}$ and $S_j$ ($j = 1, 2$) are as defined in Sec.~\ref{sec:models}.
Then $U_{\mathrm{G}} = S_{\mathrm{G}}C_{\mathrm{G}}$ defines
the evolution operator for the GGW model. 

A spectral mapping technique,
first elaborated by Szegedy \cite{Szegedy}
and further developed by several authors \cite{HKSS2014, SegawaSuzukiSMT1, SegawaSuzukiSMT2, AFSST},
clarifies the relation between the GGW
and our model.
To see this, rewriting $U_{\mathrm{G}}$ as 
\[
    U_{\mathrm{G}}= \bm{(}S_{\mathrm{G}} (\sigma_x \oplus \sigma_x) \bm{)} \times 
    \bm{(}(\sigma_x \oplus \sigma_x)C_{\mathrm{G}} \bm{)}
    =: S_{\mathrm{G}}^\prime C_{\mathrm{G}}^\prime
\]
clarifies the chiral symmetry of $U_{\mathrm{G}}$,
which stems from the fact that $S_{\mathrm{G}}^\prime$ and $C_{\mathrm{G}}^\prime$ 
are unitary involutions,
i.e.,
$a^\dagger = a^{-1} = a$ (and thus $a^2 = 1$).
See \cite{suzukiAsymptoticVelocityPositiondependent2016} for the relation between the chiral symmetry and the spectral mapping technique.
Actually, the unitarity and involutivity of $S_{\mathrm{G}}^\prime$
follow from the fact that it acts on $\mathcal{H} \oplus \mathcal{H}$
as the direct sum $S_1^\prime \oplus S_2^\prime$ of unitary involutions.
These properties for $C_{\mathrm{G}}^\prime$
are clear because one can write 
\[
    C_{\mathrm{G}}^\prime = 1 - 2\ketbra|\Phi_0><\Phi_0|,
\]
where 
$\ket|\Phi_0>$
is the transpose of
$\left(\sqrt{p}, \sqrt{p}, -\sqrt{q}, -\sqrt{q} \right) / \sqrt{2}$.

Define an operator $T_{\mathrm{G}}$ on $\ell^2(\mathbb{Z}^2)$ by
\[
    T_{\mathrm{G}}\psi = \bra<\Phi_0|S_{\mathrm{G}}^\prime \left(\psi \otimes \ket|\Phi_0>\right)
\]
for any $\psi \in \ell^2(\mathbb{Z}^2)$. 
This operator, called the discriminant, plays an important role in the spectral mapping technique.
As shown in \cite{FudaFunakawaSuzuki2017}, 
the action of $T_G$ is given by
\begin{align*}
    (T_{\mathrm{G}}\psi)(\bm{x}) &= \frac{p}{2}\bm{(}\psi(\bm{x} + \bm{e}_1) + \psi(\bm{x} - \bm{e}_1)\bm{)} \\
    & \quad + \frac{q}{2}\bm{(}\psi(\bm{x} + \bm{e}_2) + \psi(\bm{x} - \bm{e}_2)\bm{)}.
\end{align*} 
The discriminant $T_{\mathrm{G}}$ for the GGW is 
the generator of the 2D continuous-time QW (CTQW) \cite{konnoQuantumWalks2008a}
with horizontal and vertical transition probabilities $p / 2$ and $q / 2$. 
It is also commonly known as 
the (free) discrete Schr\"{o}dinger operator on $\mathbb{Z}^2$. 
Because $T_{\mathrm{G}}$ is translation invariant on $\mathbb{Z}^2$,
the Fourier transform diagonalizes
$T_{\mathrm{G}}$ in the wave vector space as
\begin{equation}
\label{def:TGhat}
    \hat{T}_{\mathrm{G}}(\bm{k}) =  p \cos k_1 + q \cos k_2, \quad \bm{k} \in \mathbb{T}^2.
\end{equation} 

The following proposition summarizes the spectral mapping technique,
which facilitates
the calculation of the spectrum of $U_{\mathrm{G}}$:
\begin{prop}[Spectral mapping] \label{prop:smt}
The absolutely continuous part of $U_{\mathrm{G}}$
is unitarily equivalent to
\begin{equation*}
    e^{i \arccos T_{\mathrm{G}}} \oplus e^{-i \arccos T_{\mathrm{G}}}
\end{equation*} 
acting on $\mathcal{H}\simeq \ell^2(\mathbb{Z}^2) \oplus \ell^2(\mathbb{Z}^2)$. 
\end{prop}
\begin{proof}
The construction developed in \cite{SegawaSuzukiSMT2} of the generator 
for chiral symmetric unitary operators implies
that 
the absolutely continuous part of $U_{\mathrm{G}}$ is unitarily equivalent
to 
$e^{i \arccos T_{\mathrm{G}}}  \oplus e^{-i\arccos T_{\mathrm{G}}}$.
\end{proof}
This proposition implies that
the dispersion relation $\omega_{\mathrm{G}}(\bm{k})$ of the GGW 
satisfies
$\cos \omega_{\mathrm{G}}(\bm{k}) = \hat{T}_{\mathrm{G}}(\bm{k})$
and the absolutely continuous part of $U_{\mathrm{G}}$ can be diagonalized as 
$e^{i \omega_{\mathrm{G}}(\bm{k})} \oplus e^{-i \omega_{\mathrm{G}}(\bm{k})}$.

To relate the GGW and our evolution $U=S_2C_2S_1C_1$,
which has been assumed to have the coins $C_j = C(a_j; 0, 0, 0)$
defined in \eqref{defeq:coin_operators},
recasting the parameters $(p, q)$ of the GGW 
as $(a, b)$ requires
that $p + q = 1$ if and only if $(a, b) \in \Delta_1$. 
\begin{thm}
\label{thm:Groverto2D2SQWs}
Let $(p, q) \in \Delta_1$.
Then,
the absolutely continuous part of $U_{\mathrm{G}}$
is unitarily equivalent to $U = S_2C_2S_1C_1$
with $C_j = C(\sqrt{p}; 0, 0, 0)$ for $j = 1, 2$.     
\end{thm}
\begin{proof}
As mentioned in the proof of Lemma \ref{lem:EFaresame},
$(a, b) \in \Delta_1$ if and only if $a_1 = a_2 = \sqrt{a}$. 
Comparing \eqref{defeq:tau} with \eqref{def:TGhat}
and applying the change of variables
${\bm{k}}^\prime=(k_1 + k_2,k_2 - k_1 + \pi)$,
we obtain the relation
\begin{align*} 
    \tau(\bm{k}) 
    & = a \cos (k_2 + k_1) + b \cos (k_2 - k_1 + \pi) 
     = T_{\mathrm{G}}({\bm{k}}^\prime),
\end{align*}
where we have used $(p, q) = (a, b)$ in the last equality. 
As shown in Sec.~\ref{sec:proofs},
the Fourier transform diagonalizes our evolution operator $U$ 
as 
$e^{i \omega(\bm{k})} \oplus  e^{-i \omega(\bm{k})}$. 
The periodicity of $\mathbb{T}^2$
ensures the existence of a unitary transformation between the wave vector spaces
that implements the change of variables 
${\bm{k}} \mapsto {\bm{k}}^\prime$. 
Therefore, 
$e^{i \arccos T_{\mathrm{G}}} \oplus e^{-i \arccos T_{\mathrm{G}}} \simeq U$.
\end{proof}
The following corollary, 
a direct consequence of Theorem \ref{thm:Groverto2D2SQWs}, 
generalizes the WLT obtained by Watabe et al. \cite{watabeLimitDistributionsTwodimensional2008} to general initial states.
To emphasize the dependence on $(p, q)$,
we use $C_{\mathrm{G}}(p, q)$ to denote
the generalized Grover coin $C_{\mathrm{G}}$ defined in \eqref{Grovercoin}. 
\begin{cor}
Let $(p, q) \in \Delta_1$ and $\bm{X}_t^{\mathrm{G}}$
be the position of the GGW with the coin operator $C_{\mathrm{G}}(p,q)$
and the initial state $\Psi^{\mathrm{G}}_0$.
Then the limiting distribution of $\bm{X}_t^{\mathrm{G}}/t$
is represented as
\[
    \left[
        \mu_{\mathrm{G}} \delta_0(v_1, v_2) + 
        w_{\mathrm{G}} (v_1, v_2) f (v_1, v_2) \right]
    dv_1dv_2,
\]
where the initial state $\Psi^{\mathrm{G}}_0$ determines the weight function $w_{\mathrm{G}}$ and the nonnegative constant $\mu_{\mathrm{G}}$. 
\end{cor}
\begin{proof}
The proof follows the standard approach developed 
in \cite{suzukiAsymptoticVelocityPositiondependent2016, SegawaSuzukiSMT2}. 
It shows that
$\mu_{\mathrm{G}}$ is the probability $P(V_1 = V_2 = 0)$
that the asymptotic velocity is zero,
and it is given by the squared norm of
the projection of $\Psi^{\mathrm{G}}_0$ to $\ker (U^2 - 1)$.
The weight function $w_{\mathrm{G}}$ is defined by \eqref{eq:w_bullet}
with the initial state $\Psi^{\mathrm{G}}_0$ replaced by the projection of $\Psi^{\mathrm{G}}_0$ to $\ker (U^2 - 1)^\perp$.
\end{proof}
Regarding prior work, 
Di Franco et al.~\cite{difrancoAlternateTwodimensionalQuantum2011} 
derived a PDF whose support is the interior of $E$ 
(defined in \eqref{ellipse}), 
while Watabe et al.~\cite{watabeLimitDistributionsTwodimensional2008} calculated the map's image, identifying this interior and four specific boundary points. 
Since the joint spectrum $\Sigma(\hat{\bm v})$ must be closed,
our analysis completes this picture by taking the closure, 
confirming $\Sigma(\hat{\bm v})$ is the entire closed ellipse $E$. 
Notably, the ${\bm k} \mapsto {\bm k}^\prime$ transformation in the proof of Theorem~\ref{thm:Groverto2D2SQWs} induces a $\pi/4$-rotation; 
the PDF in \cite{watabeLimitDistributionsTwodimensional2008} corresponds to our PDF $f$ (defined in \eqref{KonF2}) rotated by $\pi/4$.

\subsubsection{Alternate quantum walks}
Our model includes the AQW model~\cite{difrancoAlternateTwodimensionalQuantum2011}
with the coin operators
\begin{align*}
    C_1 = C_2 &= C\left( \cos \gamma; \dfrac{\pi}{2}, \dfrac{\pi}{2}, -\pi \right) 
    = \begin{pmatrix} \cos \gamma & \sin \gamma \\ \sin \gamma & -\cos \gamma \end{pmatrix}.
\end{align*}
In this case, 
$(a, b) \in \Delta_1$
since
$a = \cos^2 \gamma$, $b = \sin^2 \gamma$,
and $a + b = 1$. 
By choosing the coin operator in this way, 
all AQWs are represented by
our model in the regime $\Delta_1$.

In \cite{difrancoAlternateTwodimensionalQuantum2011},
Di Franco~et al.~demonstrate the possibility of
mimicking the GGW 
using the alternate QW. 
More precisely,
\cite[Theorem 2]{difrancoAlternateTwodimensionalQuantum2011}
states that for the initial state $\Psi_0$
with $\Psi_0(\bm{x}) = \bm{0}$ except for the origin $\bm{x} = \bm{0}$,
there exists an initial state of the GGW
such that the probability distribution of the position for 
the AQW coincides with 
that of the GGW for any time $t$. 
Theorem \ref{thm:Groverto2D2SQWs} 
generalizes this fact 
to any initial state $\Psi_0$
(even those where $\Psi_0(\bm{x}) \neq \bm{0}$ for all $\bm{x} \neq \bm{0}$)
of the AQW.

We also note the 2D2SQW model presented in \cite[Sec.~5.2]{cedzichExponentialTailEstimates2024}, defined by the evolution operator $W(k_1, k_2) = e^{ik_1\sigma_x}e^{ik_2\sigma_z}$ on $\mathbb{T}^2$. Although this model appears different from an AQW, we can show its unitary equivalence. 
By utilizing the Hadamard matrix $H$, which diagonalizes $\sigma_x$, 
we find $H W(k_1, k_2) H = e^{ik_1\sigma_z} C_1 e^{ik_2\sigma_z} C_2$ with $C_1 = C_2 = H$. 
Based on Lemma \ref{lem:equiv}, this evolution is unitarily equivalent to our model with coin operators $C_1 = C_2 = C(1/\sqrt{2}; 0, 0, 0)$. This corresponds precisely to the case $a = b = 1/2$, yielding $v_{\mathrm{max}} = a + b = 1$. 
Consequently, our general result for the $\Delta_1$ regime predicts that the joint spectrum $\Sigma(\hat{\bm{v}})$
—the ellipse $E$ defined in \eqref{ellipse}—degenerates 
into a unit circle for this specific case, consistent with the implications of the analysis in \cite{cedzichExponentialTailEstimates2024}.

The last and present subsections imply that Theorem \ref{thm:WLT'} alone recovers the known results for the 
limit distribution of QWs on the square lattice
\cite{watabeLimitDistributionsTwodimensional2008, difrancoAlternateTwodimensionalQuantum2011}. 
Notably, Theorem \ref{thm:WLT}, the central finding of this paper, 
introduces a previously unexplored class of QWs whose limit distributions are explicitly obtained. 
The 2D Konno functions derived in Theorem \ref{thm:WLT} are fundamentally different from those found in previous works,
offering a fresh perspective for the study of QWs.

\subsubsection{Generalization to higher dimensions}
As has been sketched in the Introduction, the closed form of the weight function $w$ requires properly defining the inverse functions of the two-valued group velocity $v_{\pm}(\bm{k})$. In the 1D case, this is straightforward, partitioning the wave number space into two invertible domains.

Our analytical success in the 2D case hinges on a crucial property:
the problem can be decomposed into a chain of 2D $\to$ 2D transformations. As detailed in Sec.~\ref{sec:proofs}, 
the wave vector $\bm{k} = (k_1, k_2) \in \mathbb{T}^2$ maps to an intermediate parameter space $\bm{c} = (c_1, c_2) \in [-1, 1]^2$, 
where $c_1 = \cos(k_2 + k_1)$ and $c_2 = \cos(k_2 - k_1)$.
As $\bm{k}$ moves in $\mathbb{T}^2$, 
the parameters $c_1$ and $c_2$ move independently 
within the square $[-1, 1]^2$. 
The velocity $\bm{v} = (v_1, v_2)$ is also a function of $\bm{c}$. 
This enabled us to carefully create a piecewise inverse map $\bm{k}(\bm{v})$ by dividing the shared, independent $c_1c_2$-plane.

This tractable 2D-to-2D correspondence breaks down completely in higher dimensions. For instance, in the 3D version, 3D2SQW, the corresponding $\tau(\bm{k})$ is expressed in terms of \emph{four} parameters 
$\bm{c} = (c_0, c_1, c_2, c_3)$. 
These parameters are defined by the 3D wave vector $\bm{k} = (k_1, k_2, k_3)$ as:
\begin{gather*}
	c_0 = \cos(k_1 + k_2 + k_3),\quad c_1 = \cos(k_1 - k_2 + k_3), \\
	c_2 = \cos(k_1 + k_2 - k_3),\quad c_3 = \cos(k_1 - k_2 - k_3). 
\end{gather*}
Here, the map from the 3D wave vector $\bm{k} \in \mathbb{T}^3$ 
to the parameter space $\bm{c} \in [-1, 1]^4$ is a map from three dimensions to four. 
This dimensional mismatch means the $c_i$ variables 
cannot move independently; 
they are confined to a 3D manifold embedded within the 4D hypercube.

This loss of independence makes a direct analytical inversion intractable.
Furthermore, computing the Jacobians involves analyzing homogeneous cubic polynomials in four constrained variables. 
Given these fundamental difficulties in 3D, 
it is clear that our analytical method 
for deriving the exact PDF does not generalize 
straightforwardly to three- or higher-dimensional lattices.

\section{Sketch of proofs} \label{sec:proofs}
This section outlines the proofs for our main results:
Theorem~\ref{thm:WLT} (the $v_{\mathrm{max}} < 1$ regime)
and Theorem~\ref{thm:WLT'} (the $v_{\mathrm{max}} = 1$ regime).

Our analysis is based on the framework of Grimmett et al.~\cite{grimmettWeakLimitsQuantum2004},
which defines the limiting PDF via the joint spectral measure
$d \braket <\Psi_0 | E_{\hat{\bm{v}}}(\bm{v}) \Psi_0 >$.
The core challenge in 2D is that the group velocity $\bm{v}_{\star}\colon \mathbb{T}^2 \rightarrow \Sigma(\hat{\bm{v}})$
is a highly non-injective map.
To derive the PDF, we must analytically perform the change of variables
from the wave vector $\bm{k}$ to the velocity $\bm{v}$
by rigorously inverting this map.

Our proof proceeds as follows:
First, we determine the joint spectrum $\Sigma(\hat{\bm{v}})$ (the PDF's support).
Second, we calculate the Jacobian $J$ of the velocity map,
showing its zeros (the caustics) define the support's boundary.
Third, we use the Jacobian's sign to partition $\mathbb{T}^2$
into subdomains where the map is invertible.
We then explicitly construct the inverse maps $\bm{k}(\bm{v})$.
Finally, we use these components to transform the spectral measure
from $\bm{k}$-space to $\bm{v}$-space,
yielding the final PDF $\sum_{\bullet} w_{\bullet} f_{\bullet}$.
\subsection{\texorpdfstring{Joint Spectrum $\Sigma(\hat{\bm{v}})$}{Joint Spectrum Sigma(hat v)}}
\label{subsec:jointspectrum}

The translation-invariance property
allows us to diagonalize the evolution $U$ into
the two-by-two unitary matrix-valued function $\hat{U}(\bm{k})$
in the wave vector space $\mathbb{T}^2$ and represent its spectral decomposition as 
\begin{align*} 
\hat{U} = 
    e^{i \omega} \ketbra |\omega_+><\omega_+|
    +
     e^{-i \omega} \ketbra |\omega_-><\omega_-|, 
\end{align*}
where $\omega(\bm{k}) = \arccos \tau(\bm{k})$ 
is the dispersion relation 
with $\tau(\bm{k})$ defined in \eqref{defeq:tau},
and $\ket|\omega_\star(\bm{k})>$ ($\star = \pm$) forms an ONB of eigenbasis
for $\bm{k} = (k_1, k_2) \in \mathbb{T}^2$. 
The velocity operator $\hat{\bm{v}}$ in the wave vector space is given by
the vector $\hat{\bm{v}} = (\hat{v}_1, \hat{v}_2)$ of two-by-two matrices
\begin{equation}
\label{def:velocityop}
    \hat{v}_j
    =
     \sum_{\star = \pm} v_{\star, j} \ketbra |\omega_\star><\omega_\star|,
     \quad j = 1, 2,
\end{equation}
where the functions $v_{\pm, j}$ are characterized by
the group velocities 
$\mp \nabla \omega=(v_{\pm, 1}, v_{\pm, 2})$
corresponding to the modes $\ket |\omega_\pm>$.
Note that the group velocity here is defined proportional to $\mp \nabla \omega$, corresponding to the time evolution $e^{\pm i\omega t}$.
This sign convention is consistent with the 1D QW literature
\cite{konnoQuantumWalks2008a}. 

To streamline the description, we introduce the following notation
for any wave vector $\bm{k} = (k_1, k_2)$:
\begin{align}
	l_1 &:= k_2 + k_1, & l_2 &:= k_2 - k_1, \label{defeq:l} \\
	c_1 &:= \cos l_1, & c_2 &:= \cos l_2, \label{defeq:c} \\
	s_1 &:= \sin l_1, & s_2 &:= \sin l_2, \label{defeq:s} \\
	v_1 &:= -\dfrac{as_1 + bs_2}{\sqrt{1 - \tau^2}},
	& v_2 & := -\dfrac{as_1 - bs_2}{\sqrt{1 - \tau^2}}. \label{defeq:v}
\end{align}
Here the functions $\tau$
and $v_{\star, j}$ appearing in \eqref{def:velocityop}
can be represented, respectively, as $\tau = ac_1 - bc_2$ and 
\begin{equation} \label{vpm}
    v_{\pm, j} = \pm v_j,
    \quad j = 1, 2. 
\end{equation}
The assertions (i) and (ii) of Proposition~\ref{prop:hat_v_q}
are a direct consequence of \eqref{defeq:v},
since $\hat{v}_j$ is the direct sum of the two multiplication operators by the functions $v_{\star, j}$, whose spectrum is given by the union of the ranges of $v_{\star, j}$ ($\star = \pm$) as functions.
In case of $(a, b) \in \Delta_1$,
it follows from \eqref{defeq:tau} that 
$\tau(\bm{k})^2 = 1$ at four points $(0, \pm \pi)$ and $(\pm \pi, 0)$,
but \eqref{defeq:v} can be defined almost everywhere. 

We conclude this subsection 
by stating the following theorem
that clarifies the joint spectrum $\Sigma(\hat{\bm{v}})$.
The idea of the proof 
can be found in Appendix \ref{appen:sigma_v}.
\begin{thm} \label{thm:jointspectrum}
The joint spectrum 
is given by
\[
    \Sigma(\hat{\bm{v}}) 
    = \begin{cases} 
        E_1 \cap E_2, & (a, b) \in \Delta_{(0, 1)}, \\ 
        E, & (a, b) \in \Delta_1, 
    \end{cases}
\]
where $E$ and $E_j$'s are the ellipses defined in \eqref{ellipse}
and \eqref{ellipses}. 
\end{thm}

\subsection{Jacobians of Velocity Map}
\label{sec:Jacodet}
In this section,
we calculate the absolute value of the Jacobian
for the velocity map 
${\bm v}_{\star} = (v_{\star, 1}, v_{\star, 2})$ with $\star = \pm$. 
As established in Theorem \ref{thm:jointspectrum},
these maps define 2D transformations ${\bm v}_{\star}\colon \mathbb{T}^2 \to \Sigma(\hat{\bm{v}})$.
From \eqref{vpm} and multi-linearity of determinants, 
the Jacobians of $v_{\star}$ are identical. 
We therefore denote this common determinant by $J$,
which can be written as
\begin{equation*} 
J = \det \left( \nabla \bm{v} \right) 
\end{equation*}
with $\bm{v} = (v_1, v_2)$ defined in \eqref{defeq:v}.

The remainder of this subsection
is devoted to a sketch of the calculation of $J$ 
and the absolute value of its inverse, $|J|^{-1}$,
as functions of $\bm{v} = (v_1, v_2)$.
Since the full derivation is algebraically intensive,
we highlight only the principal steps and 
leave the details to the reader.
\subsubsection{\texorpdfstring{
Quadratic form $\varphi({\bm c})$ for the Jacobian $J$
}{Quadratic form varphi(c1, c2) for the Jacobian J}}
We introduce additional notation that makes
calculations more tractable:
\begin{equation}
\label{defeq:taupm}
    \tau_{\pm} := ac_1 \pm bc_2, \quad \sigma_{\pm} := as_1 \pm bs_2,
\end{equation}
where $c_i, \ s_i$ 
are defined in Eqs.~\eqref{defeq:c}--\eqref{defeq:s}
and we observe that $\tau = \tau_-$.
A key advantage of this notation is that it provides compact expressions for the gradients:
\[ 
    \nabla \sigma_{\pm} = (\tau_{\mp}, \tau_{\pm}),
    \quad \nabla \tau_{\pm} = -(\sigma_{\mp}, \sigma_{\pm}).
\]
Consequently, 
the velocity $\bm{v}$ 
and its derivative can be expressed
solely in terms of $\tau_{\pm}$ and $\sigma_{\pm}$;
The following proposition, 
obtained by using such relations,
serves as the starting point
of our analysis.
\begin{prop}
\label{prop:reprJacob}
The Jacobian is represented
in terms of ${\bm c} = (c_1, c_2)$ and $\tau$ as
\begin{equation} \label{repr:Jacobdet}
    J = - \dfrac{4ab \varphi({\bm c}) }{(1 - \tau^2)^2} ,
\end{equation}
where the function $\varphi$ is a quadratic form given by 
\begin{equation} \label{def:varphi}
    \varphi({\bm c}) = ab c_1^2 + (1 - a^2 - b^2)c_1c_2 + ab c_2^2.
\end{equation}
\end{prop}
Observe from \eqref{repr:Jacobdet} that
the quadratic form $\varphi$ is of central importance,
as its zeros correspond to the vanishing of $J$.
These zeros, therefore, define the boundaries
of our domain separation,
which physically represent the caustics of the group velocity map,
or equivalently, the locus of the divergent effective mass.

\subsubsection{\texorpdfstring{Quadratic equation for $1 - \tau^2$}{Quadratic equation for 1 - tau2}}
As a result of Proposition \ref{prop:reprJacob},
the Jacobian $J$ has been expressed
in terms of $c_i$ and $1 - \tau^2$.
Here, we represent $1 - \tau^2$ in terms of the velocity $\bm{v}$.

To this end, squaring both sides of \eqref{defeq:tau} provides
\begin{equation}
    \tau^2 - a^2c_1^2 - b^2c_2^2 = -2abc_1c_2. \label{eq:tau2}
\end{equation}
Squaring both sides of \eqref{eq:tau2} again, 
and then substituting the following
(obtained from \eqref{defeq:v})
\begin{equation}
    c_i^2= 
    \begin{cases}
        1 - {(1 - \tau^2)(v_1 + v_2)^2} / (4a^2), & i = 1, \\ 
        1 - {(1 - \tau^2)(v_1 - v_2)^2} / (4b^2), & i = 2
    \end{cases}
    \label{eq:c_v}
\end{equation}
eliminates the functions $c_j^2$ and finds
that $T := 1 - \tau^2$ satisfies the following quadratic equation:
\begin{equation} 
    A T^2 - 4abF_0 T + B = 0, \label{eq:EQ_of_1-tau2}
\end{equation}
where $F_0$ is defined in Eq.~\eqref{eq:B_with_v}
and
\begin{align*}
    & A = (1 - v_1^2)(1 - v_2^2), \\
    & B = \left[1 - (a + b)^2\right]\left[1 - (a - b)^2\right].
\end{align*} 
\begin{prop}
\label{prop:quadraticeq}
Let $D$ be the discriminant
of the quadratic equation \eqref{eq:EQ_of_1-tau2}. 
Then:
\begin{itemize}
    \item[(1)] The reduced discriminant 
        $D^\prime := D/4 = (2abF_0)^2 - AB$ is equal to $(2ab)^2 F_1 F_2$:
        \[(2abF_0)^2 - AB = (2ab)^2 F_1 F_2 \]
    \item[(2)] The solution $T= 1-\tau^2$ of 
        the quadratic equation \eqref{eq:EQ_of_1-tau2} is given by
        \begin{equation}
            1 - \tau^2 
                = 2ab \left[\dfrac{ F_0 \pm \sqrt{F_1F_2}}
                {(1 - v_1^2)(1 - v_2^2)} \right].
                \label{eq:1-tau2}
        \end{equation}
\end{itemize}
\end{prop}
We omit the proof, but offer some brief remarks:
\begin{itemize}
    \item While (2) follows straightforwardly from (1)
        by applying the quadratic formula to \eqref{eq:EQ_of_1-tau2},
        the verification of statement (1) is \emph{far from trivial} 
        and involves a tedious algebraic calculation of $(2abF_0)^2 - AB$.
    \item The RHS of \eqref{eq:1-tau2} 
        is a two-valued function of $\bm{v}$,
        an ambiguity that stems from the non-injectivity of $\bm{v} = \bm{v}(\bm{k})$.
        Resolving this requires choosing a specific branch
        for the inverse function $\bm{k}(\bm{v})$,
        which is achieved by restricting the domain $\mathbb{T}^2$
        to a region where the mapping is injective.
        The choice of branch determines the sign $\pm$ of the RHS.
\end{itemize}

\subsubsection{Caustics and Jacobian}

For $(a, b) \in \Delta_{(0, 1)}$,
Propositions \ref{prop:reprJacob} and \ref{prop:quadraticeq}
allow us to eliminate $c_1$ and $c_2$ from the quadratic form $\varphi(\bm{c})$ defined in \eqref{def:varphi} 
and represent the Jacobian $J$
as a function of $v$. 
Actually, Eqs.~\eqref{eq:tau2}--\eqref{eq:c_v}
eliminate $c_1$ and $c_2$, 
and then using Eqs.~\eqref{eq:EQ_of_1-tau2}--\eqref{eq:1-tau2}
establishes
\begin{align*}
    - 2ab \varphi(\bm{c})
    & = B - 2ab F_0 T \\
    & = T (2ab F_0  - A T) = \mp 2ab T \sqrt{F_1F_2}.
\end{align*}
Substituting this into the RHS of \eqref{repr:Jacobdet}
provides
\begin{equation} \label{repr:Jacobdet2}
    J = \dfrac{\mp 4ab\sqrt{F_1F_2}}{1 - \tau^2}.
\end{equation}
Combining this with \eqref{eq:1-tau2}
and taking into consideration the second remark on the sign $\pm$
corresponds to the branches of the inverse function $\bm{k} = \bm{k}(\bm{v})$,
we can represent the reciprocal of the Jacobian $J$
in terms of the velocity $\bm{v} = (v_1, v_2)$ as
\begin{equation}
\label{eq:recipJaco}
    1/J = \begin{cases}
        - \pi^2 f_+, & J < 0, \\
        + \pi^2 f_-, & J > 0,
    \end{cases} 
\end{equation}
where the functions $f_{\pm}$ are the 2D Konno functions
defined in \eqref{eq:fpm}.
For notational simplicity,
we write the absolute values of the RHS of \eqref{eq:recipJaco}
as $|J|_{+1}^{-1}$ for $J < 0$ and $|J|_{-1}^{-1}$ for $J > 0$,
which are identical, from a standard result in bivariate calculus, 
with the absolute value of the Jacobian
of an inverse function $\bm{k} = \bm{k}(\bm{v})$.

Moreover, combining Theorem \ref{thm:jointspectrum}
with Eqs.~\eqref{repr:Jacobdet} and \eqref{repr:Jacobdet2}
provides that
for $\bm{k} \in \mathbb{T}^2$ and $\bm{v} = \bm{v}_{\star}(\bm{k})$,
the following are equivalent:
\begin{itemize}
    \item[(a)] The Jacobian $J$ vanishes, i.e., $J(\bm{k}) = 0$.
    \item[(b)] The quadratic form $\varphi$ has zeros, 
    i.e., $\varphi(\bm{c}) = 0$.
    \item[(c)] $\bm{v}$ lies on 
    $\partial (E_1 \cap  E_2)$,
 i.e., $F_1(\bm{v})F_2(\bm{v}) = 0$ and $F_j(\bm{v})\geq0$. 
    \item[(d)] $\bm{v}$ lies on the boundary of $\Sigma({\hat{\bm{v}}})$,
    i.e., $\bm{v} \in \partial \Sigma({\hat{\bm{v}}})$.
\end{itemize}

For $(a, b) \in \Delta_1$
that is equivalent to that of $B = 0$, 
Proposition \ref{prop:quadraticeq} dictates
\begin{equation*}
    1 - \tau^2 = 0 \quad \text{or} \quad
    \dfrac{4abF_0}{(1-v_1^2)(1-v_2^2)}.
\end{equation*}
However, the former situation $1 - \tau^2 = 0$ happens only at 
one of the four points ${\bm k}=(\pm \pi, 0), \ (0, \pm \pi)$
where $\tau({\bm k})^2 = 1$.
Otherwise, an argument similar to above
provides $-2ab\varphi(\bm{c}) = -2abF_0 T$.
Substituting this into \eqref{repr:Jacobdet},
we obtain $|J|^{-1} = \pi^2 f$,
where the function $f$ is defined in \eqref{KonF2}
and $|J|^{-1}$ simply denotes the reciprocal of $|J|$,
since no branch exists for this case. 

In summary, we arrive at the following:
\begin{thm}[Reciprocal of Jacobian] \label{thm:Jacobinv}
Let $\Sing (J)$ 
be the singular set of the velocity map,
i.e., the zeros of the Jacobian:
$\Sing (J)=\{ \bm{k} \in \mathbb{T}^2 \mid J(\bm{k}) = 0\}$.
\begin{itemize}
    \item[(1)] Let $(a, b) \in \Delta_{(0, 1)}$.
    Then $|J|_{\pm 1}^{-1} = \pi^2 f_\pm$.
    The limit PDF forms a caustic, diverging on the image of the singular set, which coincides with the boundary of the support:
    $\bm{v}(\Sing (J))
        = \partial \Sigma(\hat{\bm{v}})$.
    \item[(2)] Let $(a, b) \in \Delta_1$. 
    Then $|J|^{-1} = \pi^2 f$. 
    The limit PDF diverges on the image of the singular set, which, in this case, is the discrete four-point set:
    $\bm{v}(\Sing (J)) 
        = \{ \pm (1, a - b), 
            \pm (a - b, 1) 
            \}$. 
\end{itemize}
\end{thm}

\subsection{Domain separations}
\label{subsec:domainseparation}
The separation of the domain for the velocity map $\bm{v} = \bm{v}_{\star}(\bm{k})$,
required to properly define its inverse, is not merely a technical procedure.  
It is fundamentally the problem of identifying the physical region of finite effective mass, distinct from the caustics where it diverges.
To implement this separation, we proceed several steps.
\begin{figure}[ht]
    \includegraphics[width=6.5cm]{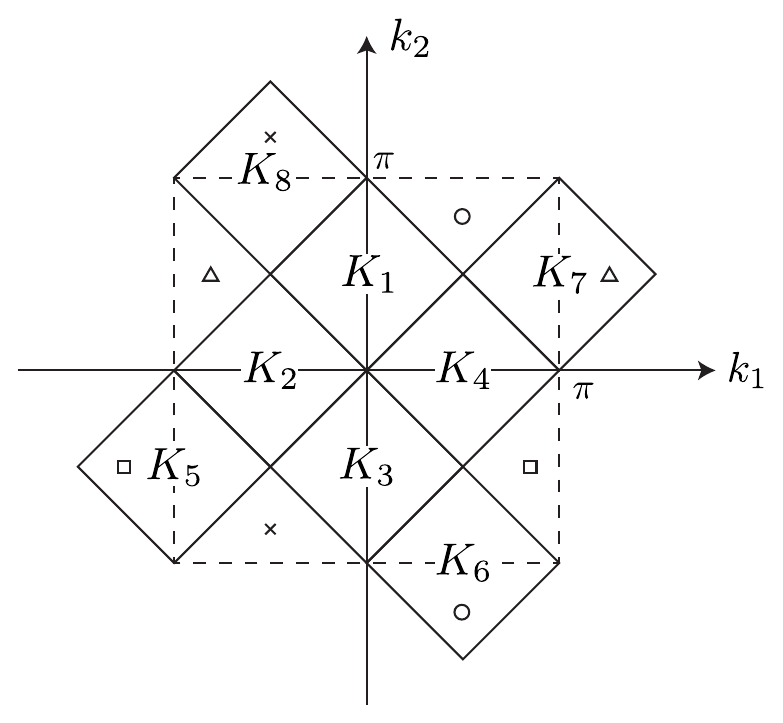}
    \caption{
        The ``windmill'' representation of the wave vector space $\mathbb{T}^2$, partitioned into eight regions $K_n$ ($n = 1, \dots, 8$). This structure is a deformation of the standard square domain (indicated by dashed lines), constructed by ``cutting and pasting'' regions based on the translational invariance of the torus. The symbols illustrate this identification: for example, the $\triangle$-marked region (top-left of the dashed-line box) is translated by $(+2\pi, 0)$ to form region $K_7$. Similarly, the $\bigcirc$-marked region (top-right of the dashed-line box) is translated by $(0, -2\pi)$ to form region $K_6$.
        This partition ensures that the signs of $s_1$ and $s_2$ by \eqref{defeq:s} are constant values within each region $K_n$ (see TABLE \ref{tbl:sgn_s1_s2}). All regions include their boundaries, which
        do not contribute to the integrals.
    } \label{fig:only_K}
\end{figure}
\subsubsection{Separating Windmill to eight domains \texorpdfstring{$K_n$}{Kn}}
The periodicity of the wave vector space $\mathbb{T}^2$
allows us to identify it with a ``windmill-like'' region,
which we then partition it into eight coarse domains 
$K_n$ ($n = 1, 2, \dots, 8$), as shown in FIG.~\ref{fig:only_K}.
Within the windmill-like domain, 
the boundaries of each domain $K_n$
are parallel to the rotated coordinate axes
$l_1$ and $l_2$ (defined in \eqref{defeq:l}).
This alignment uniquely determines the signs of 
$s_i$ ($i = 1, 2$) in each region $K_n$, 
as indicated in TABLE~\ref{tbl:sgn_s1_s2}.
\begin{table}[ht]
	\caption{Signs of $s_1$ and $s_2$} \label{tbl:sgn_s1_s2}
	\begin{tabular}{c|cc}
		$n$   & $\sgn(s_1)$ & $\sgn(s_2)$ \\ \hline
		$1,5$ & $+1$ & $+1$ \\
		$2,6$ & $-1$ & $+1$ \\
		$3,7$ & $-1$ & $-1$ \\
		$4,8$ & $+1$ & $-1$ \\ \hline
	\end{tabular}
\end{table}
\begin{figure}[ht] 
    \includegraphics[width=7cm]{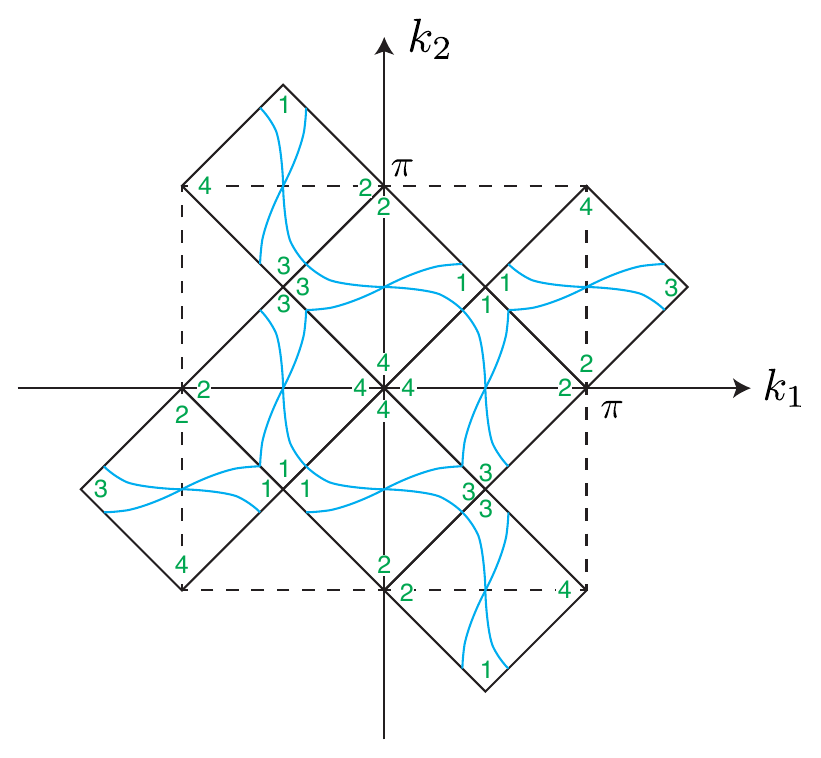}
    \caption{
        The finer partition of each region $K_n$ (from FIG.~\ref{fig:only_K}) into four subdomains $K_n^{(m)}$ ($m = 1, 2, 3, 4$).
        The blue curves defining this separation are the pullbacks 
        onto the $k_1k_2$-plane of the lines $c_2 = j_{\pm}c_1$.
        These lines, which represent the zeros of the quadratic form $\varphi({\bm c})$ (i.e., that of the Jacobian), are shown 
        in FIG.~\ref{fig:C_m}.
        The green numerical label $m$ assigned to each subdomain $K_n^{(m)}$ corresponds to the subscript of the region $C^{(m)}$ in FIG.~\ref{fig:C_m}.
    } \label{fig:windmill_waved_lines}
\end{figure}

\subsubsection{\texorpdfstring{Partition $K_n$ into four subdomains $K_n^{(m)}$}{Partition Kn into four subdomains Knm}} \label{subsubsec:Knm}
In order that the restrictions $\bm{v}_{\star}|_{K_n^{(m)}}$ 
of the velocity map are invertible,
we further partition $K_n$ into four subdomains,
$K_n^{(m)}$ ($m = 1, 2, 3, 4$),
which are illustrated in FIG.~\ref{fig:windmill_waved_lines}.
The key to define such a finer separation
is mapping each region $K_n$ onto 
the square region $[-1, 1]^2$ in the $c_1c_2$-plane
and defining the separation in the $c_1c_2$-plane.
This allow us to avoid the non-injectivity
originating from the trigonometric functions
$c_i$ and $s_i$.
Then the quadratic form $\varphi({\bm c})$ successfully 
determines the sign of the Jacobian $J$
in the $c_1c_2$-plane as stated in the following proposition
(we give the proof in Appendix \ref{pr:prop:boundaryCm}):
\begin{figure}[ht] 
    \includegraphics[width=5.5cm]{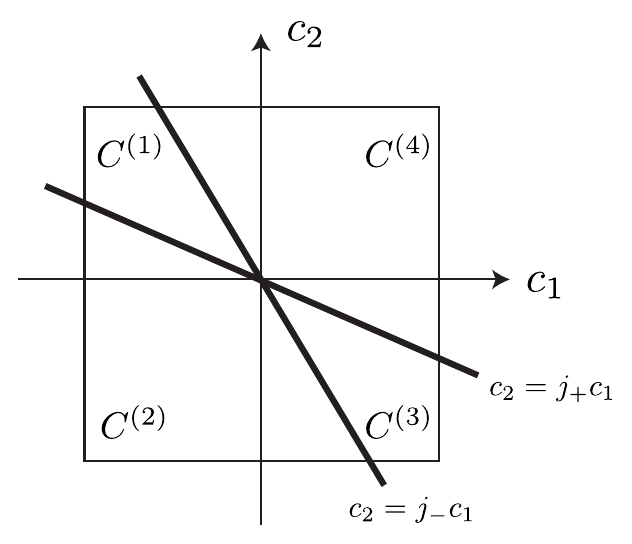}
    \caption{
        The separation of the $c_1c_2$-plane into the four regions $C^{(m)}$ ($m=1, 2,3, 4$) by the lines $c_2 = j_{\pm}c_1$. These regions correspond to the subdomains $K_n^{(m)}$ shown in FIG.~\ref{fig:windmill_waved_lines}. The constants $j_{\pm}$ are slopes that satisfy the inequalities $j_- < -1 < j_+ < 0$. 
        The degenerate case $j_{+} = j_{-} = -1$ occurs if and only if $v_{\mathrm{max}} = 1$. In this case, the regions $C^{(1)}$ and $C^{(3)}$ collapse. 
    }\label{fig:C_m}
\end{figure}
\begin{table}
	\caption{Inverse function $\bm{l} = \bm{l}(\bm{c})$} \label{tbl:C_inv}
	\begin{ruledtabular}
		\begin{tabular}{cl|cl}
			$n$ & $\bm{l} = \bm{l}(\bm{c})$ & $n$ & $\bm{l} = \bm{l}(\bm{c})$\\ \hline
			$1$ & $\begin{pmatrix} l_1 \\ l_2 \end{pmatrix} = \begin{pmatrix} \Arccos c_1 \\ \Arccos c_2 \end{pmatrix}$ 
            & $5$ & $\begin{pmatrix} l_1 \\ l_2 \end{pmatrix} = \begin{pmatrix} \Arccos c_1 - 2\pi \\ \Arccos c_2 \end{pmatrix}$ \\
			$2$ & $\begin{pmatrix} l_1 \\ l_2 \end{pmatrix} = \begin{pmatrix} -\Arccos c_1 \\ \Arccos c_2 \end{pmatrix}$ 
            & $6$ & $\begin{pmatrix} l_1 \\ l_2 \end{pmatrix} = \begin{pmatrix} -\Arccos c_1 \\ \Arccos c_2 - 2\pi \end{pmatrix}$ \\
			$3$ & $\begin{pmatrix} l_1 \\ l_2 \end{pmatrix} = \begin{pmatrix} -\Arccos c_1 \\ -\Arccos c_2 \end{pmatrix}$ 
            & $7$ & $\begin{pmatrix} l_1 \\ l_2 \end{pmatrix} = \begin{pmatrix} -\Arccos c_1 + 2\pi \\ -\Arccos c_2 \end{pmatrix}$ \\
			$4$ & $\begin{pmatrix} l_1 \\ l_2 \end{pmatrix} = \begin{pmatrix} \Arccos c_1 \\ -\Arccos c_2 \end{pmatrix}$
            & $8$ & $\begin{pmatrix} l_1 \\ l_2 \end{pmatrix} = \begin{pmatrix} \Arccos c_1 \\ -\Arccos c_2 + 2\pi \end{pmatrix}$ 
		\end{tabular}
	\end{ruledtabular}
\end{table}

\begin{prop}[Branches of Jacobian] \label{prop:boundaryCm}
The sign of the Jacobian $J$ is determined by the zeros of the quadratic form $\varphi({\bm c})$ (defined in Eq.~\eqref{def:varphi}). These zeros form lines $c_2 = j_\pm c_1$ that partition the $c_1c_2$-plane into regions $C^{(m)}$, as depicted in FIG.~\ref{fig:C_m}. The specific partitioning depends on the parameter regime:
\begin{itemize}
    \item[(1)] In the $\Delta_{(0, 1)}$ regime:
    There exist unique real constants $j_\pm$ satisfying $j_- = 1 / j_+$ and $-1 < j_+ < 0$. These lines divide the square $[-1, 1]^2$ into four regions $C^{(m)}$ ($m=1, 2, 3, 4$). On each region $C^{(m)}$ (excluding the boundary), the reciprocal of the Jacobian takes a specific branch: $|J|^{-1} = |J|^{-1}_{(-1)^m}$ .
    \item[(2)] In the $\Delta_1$ regime:
    This regime corresponds to the degenerate case $j_+ = j_- = -1$. The line $c_2 = -c_1$ divides the square $[-1, 1]^2$ into two non-empty regions $C^{(m)}$ ($m=2, 4$) . On these regions (excluding the boundary), the reciprocal of the Jacobian is $|J|^{-1}=|J|^{-1}_{+1}$ .
\end{itemize}
\end{prop}
Mapping $C^{(m)}$ back to the original wave vector space
(see TABLE \ref{tbl:C_inv} for the construction)
provides the desired separations $K_n^{(m)}$
as illustrated on the windmill in FIG.~\ref{fig:windmill_waved_lines}.
The boundaries separating these subdomains $K_n^{(m)}$
are the caustics stated in (1) of Theorem~\ref{thm:Jacobinv}.

\subsubsection{Decomposing \texorpdfstring{$K_n^{(m)}$}{Knm} into four parts \texorpdfstring{$K_{n,n^\prime}^{(m)}$}{Knn'm}}
In Sec.~\ref{subsubsec:Knm},
we successfully determined 
the invertible domains of the velocity map $\bm{v} = \bm{v}_{\star}$
($\star = \pm$).
In our construction, 
the Jacobian of $\bm{v}_{\star}$ has the same sign
on each subdomains $K_n^{(m)}$. 
The inverse function theorem
then states that the restriction $\bm{v}_{\star}|_{K_n^{(m)}}$ is injective
and therefore invertible. 

However, establishing invertibility is not sufficient.
To obtain the closed form of the limiting PDF, 
we must determine $\sgn c_i$,
since $v_{\star, j}$ and $\ket |\omega_{\star} >$
depend not only on $s_i$ but also on $c_i$. 
This demands a further separation
of each subdomain $K_n^{(m)}$.
Fortunately, this task is not complicated.
We can again leverage the $c_1c_2$-plane,
where the signs of $c_i$ are naturally determined.
In fact, each $n^\prime$-th quadrant $n^\prime = 1, 2, 3, 4$
of this plane corresponds to unique set of signs
$\bm{(}\sgn (c_1), \sgn (c_2)\bm{)}$, as summarized in TABLE~\ref{tbl:sgn_c1_c2}.
\begin{table}[ht]
	\caption{Signs of $c_1$ and $c_2$} \label{tbl:sgn_c1_c2}
		\begin{tabular}{c|cc}
			$n'$ & $\sgn(c_1)$ & $\sgn(c_2)$ \\ \hline
			$1$ & $+1$ & $+1$ \\
			$2$ & $-1$ & $+1$ \\
			$3$ & $-1$ & $-1$ \\
			$4$ & $+1$ & $-1$ \\ \hline
		\end{tabular}
\end{table}

We thus define a new, finer partition $K_{n,n^\prime}^{(m)}$
by taking the intersection of $C^{(m)}$
and the $n^\prime$-th quadrant in the $c_1c_2$-plane,
and pulling this intersection back to 
the corresponding domain $K_n^{(m)}$. 
Although quadrants are typically defined as open sets (excluding the coordinate axes), we explicitly take the closure of their intersection with $C^{(m)}$.
This procedure ensures that the boundaries
are properly included in the pullback.

Note that for certain combinations of $n^\prime$ and $m$,
this intersection in the $c_1c_2$-plane may be empty
(e.g., $(n^\prime, m)=(1,1)$ suggested by FIG.~\ref{fig:C_m}).
In such cases, the corresponding $K_{n,n^\prime}^{(m)}$
is simply considered to be the empty set.
TABLE \ref{tab:emptyset} provides a complete classification, summarizing which subdomains $K_{n,n^\prime}^{(m)}$
are non-empty and which are empty ($\emptyset$). 
\begin{table}[ht]
    \caption{Classification of subdomains $K_{n,n^\prime}^{(m)}$ as empty ($\emptyset$) or non-empty. The index $n^\prime$ corresponds to the quadrant in the $c_1c_2$-plane, and $m$ corresponds to the sign of the Jacobian.}
    \label{tab:emptyset}
    \begin{ruledtabular}
        \begin{tabular}{c|cccc}
            & $m=1$ & $m=2$ & $m=3$ & $m=4$ \\ \hline
            $n^\prime=1$ & $\emptyset$ & $\emptyset$ & $\emptyset$ & $K_{n, 1}^{(4)}$ \\
            $n^\prime=2$ & $K_{n, 2}^{(1)}$ & $K_{n, 2}^{(2)}$ & $\emptyset$ & $K_{n, 2}^{(4)}$ \\
            $n^\prime=3$ & $\emptyset$ & $K_{n, 3}^{(2)}$ & $\emptyset$ & $\emptyset$ \\
            $n^\prime=4$ & $\emptyset$ & $K_{n, 4}^{(2)}$ & $K_{n, 4}^{(3)}$ & $K_{n, 4}^{(4)}$ \\
        \end{tabular}
    \end{ruledtabular}
\end{table}
With this final partition, the signs of $c_i$
are uniquely fixed within each non-empty subdomain
$K_{n,n^\prime}^{(m)}$ (excluding the boundary). 
 As summarized in TABLE \ref{tab:emptyset}, 
 whether a given subdomain is empty depends
 by construction only on the combination of $n^\prime$
 (determining $\sgn c_i$) 
 and $m$ (corresponding to the Jacobian's sign), 
 and not on the coarse index $n$ (determining $\sgn(s_i)$).
 
This partition is now complete,
and the set of all non-empty subdomains $K_{n,n^\prime}^{(m)}$ 
collectively covers the entire wave vector space $\mathbb{T}^2$
(identical with the ``windmill'' region).
Consequently, their images under the velocity map $v$
fully cover the joint spectrum $\Sigma(\hat{\bm{v}})$.
More precisely, as the geometric arrangement of the velocity map
and TABLE \ref{tab:emptyset} show, 
the union of the images 
$\bm{v}\bm{(}K_{n,n^\prime}^{(m)}\bm{)}$
for a \emph{single} index $m$ (corresponding to a specific Jacobian sign) 
already covers the entire spectrum:
\begin{thm}[Multi-Covers of $\Sigma(\hat{\bm{v}})$]
\label{thm:multicoveredjoint}
For each $m=1, 2, 3, 4$, the joint spectrum $\Sigma(\hat{\bm{v}})$ 
can be decomposed into the following unions,
where the images $\bm{v}(K_{n,n^\prime}^{(m)})$ are
mutually disjoint excluding their boundary:
\begin{itemize}
    \item[(1)] $\displaystyle 
        \bigcup_{\substack{n=1,2,3,4}} \bm{v}\left(K_{n,2}^{(1)}\right)
        \ = \ \bigcup_{\substack{n=5,6,7,8}} \bm{v}\left(K_{n,2}^{(1)}\right) 
        = \Sigma(\hat{\bm{v}})$. 

    \item[(2)] $\displaystyle \bigcup_{\substack{n = 1, 2, 3, 4 \\ n' = 2, 3, 4}} \bm{v}\left(K_{n, n'}^{(2)}\right)
        =\bigcup_{\substack{n = 5, 6, 7, 8 \\ n' = 2, 3, 4}} \bm{v}\left(K_{n, n'}^{(2)}\right)
        = \Sigma(\hat{\bm{v}})$.
    
    \item[(3)] $\displaystyle 
        \bigcup_{\substack{n=1, 2, 3, 4}} \bm{v}\left(K_{n, 4}^{(3)}\right)
        \ = \ \bigcup_{\substack{n = 5, 6, 7, 8}} \bm{v}\left(K_{n, 4}^{(3)}\right)
        = \Sigma(\hat{\bm{v}})$.
    
    \item[(4)] $\displaystyle
        \bigcup_{\substack{n = 1, 2, 3, 4 \\ n' = 1, 2, 4}} \bm{v}\left(K_{n, n'}^{(4)}\right)
        = \bigcup_{\substack{n = 5, 6, 7, 8 \\ n' = 1, 2, 4}} \bm{v}\left(K_{n, n'}^{(4)}\right)
        = \Sigma(\hat{\bm{v}})$.
\end{itemize}
\end{thm}
This multi-covered structure of the joint spectrum
is the essential geometric feature that allows for the decomposition of the limiting PDF.

We omit a formal proof, but the key insight is as follows.
The union of domains for $n \in \{1, 2, 3, 4\}$ 
(or, alternatively, $n \in \{5, 6, 7, 8\}$) covers all four sign combinations of $(s_1, s_2)$ (TABLE \ref{tbl:sgn_s1_s2}). 
Since the $\pi / 4$-rotated velocity map $\tilde{v}$ 
maps the $(s_1, s_2)$ space to the joint spectrum $\Sigma(\hat{\bm{v}})$ (which is the elliptic intersection $\tilde{E}_1 \cap \tilde{E}_2$ 
as seen in Sec.~\ref{subsec:interpretation}), 
we need only verify the mapping (given by \eqref{defeq:v}) 
from the first quadrant, 
$(|s_1|, |s_2|)$, to the first quadrant of $\Sigma(\hat{\bm{v}})$, 
for each $m$. 
This is confirmed by showing that the boundaries of the domain $C^{(m)}$ map precisely onto the boundaries of the spectrum. 
For instance, for $m = 1$, the boundaries $c_2 = j_{\pm} c_1$ 
(see FIG. \ref{fig:C_m}) map to the elliptic arcs of $\partial\Sigma(\hat{\bm{v}})$, while the boundaries $c_i = \pm 1$ (implying $s_j = 0$ and thus $\tilde{v}_j = 0$) 
map to the $\tilde{v}$-axes.

\subsection{Inverse Velocity Maps}
\label{sec:inversevelocitymap}
We are now ready to construct the inverse velocity map $\bm{k} = \bm{k}(\bm{v})$. 
We first consider the case of $(a, b) \in \Delta_{(0, 1)}$. 
The starting point is Eqs.~\eqref{eq:c_v} and \eqref{eq:1-tau2}, which together express $|c_i|$ as a function of $\bm{v} = (v_1, v_2)$:
\begin{align}
    |c_1| &= \left[1 - \frac{b(v_1 + v_2)^2 (F_0 + (-1)^m \sqrt{F_1F_2})}{2a (1 - v_1^2)(1 - v_2^2)}\right]^{1/2}, \label{eq:c1_ex} \\
    |c_2| &= \left[1 - \frac{a(v_1 - v_2)^2 (F_0 + (-1)^m \sqrt{F_1F_2})}{2b (1 - v_1^2)(1 - v_2^2)}\right]^{1/2}, \label{eq:c2_ex}
\end{align}
where the sign $(-1)^m$ in the RHS of the above equations
corresponds to the branch of $1 - \tau^2$ in Eq.~\eqref{eq:1-tau2}
that is identical to the branch $|J|_{(-1)^m}$
for the Jacobian, as seen in Sec.~\ref{subsec:domainseparation}.
Our task is to construct the explicit inverse $\bm{k} = \bm{k}(\bm{v})$ 
for each of the ``covers'' of the joint spectrum $\Sigma(\hat{\bm{v}})$ established in Theorem \ref{thm:multicoveredjoint}.

We illustrate this procedure for an arbitrary point $\bm{v}$ belonging to one of the disjoint images. Assume $\bm{v}$ lies in the interior of $\bm{v}\bm{(}K_{3, 2}^{(1)}\bm{)}$, which corresponds to the first decomposition ($n \in \{1, 2, 3, 4\}$) in Theorem \ref{thm:multicoveredjoint} (1). We can safely ignore the boundaries, as they do not contribute to the final integral.

The multi-index $(n, n', m) = (3, 2, 1)$ provides all the information required to uniquely invert the map:

\begin{enumerate}
    \item The index $m = 1$ specifies the correct branch
    for Eqs.~\eqref{eq:c1_ex}--\eqref{eq:c2_ex},
    which corresponds to the sign $(-1)^1 = -1$ in the numerators
    of the RHS.
    
    \item The index $n' = 2$ determines the signs of $c_i$ via TABLE \ref{tbl:sgn_c1_c2} as $\bm{(}\sgn (c_1), \sgn (c_2)\bm{)} = (-1, +1)$.
\end{enumerate}

Combining these, ${\bm c}(\bm{v}) = (c_1(\bm{v}), c_2(\bm{v}))$ is uniquely determined as a function of $\bm{v}$:
\begin{align*}
    c_1(\bm{v}) &= -\left[1 - \frac{b(v_1 + v_2)^2 (F_0 - \sqrt{F_1F_2})}{2a (1 - v_1^2)(1 - v_2^2)}\right]^{1/2}, \\ 
    c_2(\bm{v}) &= +\left[1 - \frac{a(v_1 - v_2)^2 (F_0 - \sqrt{F_1F_2})}{2b (1 - v_1^2)(1 - v_2^2)}\right]^{1/2}.
\end{align*}
Finally, the index $n = 3$
selects the unique inverse branch $\bm{l} = \bm{l}(\bm{c})$ from TABLE \ref{tbl:C_inv}. 
For $n = 3$, this is 
$(l_1, l_2) = (-\Arccos c_1, -\Arccos c_2)$.
From Eq.~\eqref{defeq:l}, the inverse map $\bm{k} = \bm{k}(\bm{v})$ for $\bm{v} \in \bm{v}(K_{3, 2}^{(1)})$ is thus explicitly given by:
\[ \bm{k}(\bm{v}) 
= \frac{1}{2} \begin{pmatrix} 1 & -1 \\ 1 & 1 \end{pmatrix} \begin{pmatrix} -\Arccos c_1(\bm{v}) \\ -\Arccos c_2(\bm{v}) \end{pmatrix}. \]
This procedure can be applied similarly to all other non-empty subdomains $K_{n,n^\prime}^{(m)}$.

We next consider the case of $(a, b) \in \Delta_1$. The procedure is simpler, as this regime corresponds to the degenerate case $j_+ = j_- = -1$ (see FIG.~\ref{fig:C_m}). Consequently, the domains $C^{(1)}$ and $C^{(3)}$ collapse to measure zero, and we only need to consider the $m = 2$ and $m = 4$ branches.
Lemma \ref{lem:EFaresame} states that in this regime, $F_0 = F_1 = F_2 = F$. This significantly simplifies the expressions for $|c_i|$ in Eqs.~\eqref{eq:c1_ex} and \eqref{eq:c2_ex}. The inverse map $\bm{k}(\bm{v})$ is then constructed similarly, but only for the non-empty domains $K_{n, n'}^{(2)}$ and $K_{n, n'}^{(4)}$, which correspond to the single 2D function $f$
defined in \eqref{KonF2}.

\subsection{Derivation of the Limiting PDF}
\label{sec:DerivationPDF}
We now arrive at the final step:
the derivation of the limiting PDF, 
$\rho(\bm{v})$. 
As established in Sec.~\ref{sec:main_theorem}, 
our goal is to find the density $\rho(\bm{v})$ of the joint spectral measure $\braket <\Psi_0 | E_{\hat{\bm{v}}}(\bm{v}) \Psi_0 >$
for the velocity operator associated with the initial state $\Psi_0$.
As mentioned in Sec.~\ref{sec:main_theorem},
the function $\rho$ is formally represented as 
$\rho(\bm{v}) = |\braket < v_1, v_2|\Psi_0 >|^2$.

The standard method to identify this density $\rho(\bm{v})$ is to require that the following equality holds for any bounded continuous function $g(\bm{v}) = g(v_1, v_2)$:
\begin{align}
    \int_{\Sigma(\hat{\bm{v}})} g(\bm{v}) \rho(\bm{v})\, d\bm{v} 
    & = \int_{\Sigma(\hat{\bm{v}})} g(\bm{v})\, 
    d \braket < \Psi_0 | E_{\hat{\bm{v}}}(\bm{v}) \Psi_0>.
    \label{eq:spectral_integral_k}
\end{align}
Moving the integral to the wave vector space
(via the Fourier transform),
in which the velocity operator $\hat{\bm{v}}$ is diagonal,
yields:
\begin{align*} \eqref{eq:spectral_integral_k}
    & = \sum_{\star = \pm} \int_{\mathbb{T}^2} g\bm{(}\bm{v}_\star(\bm{k})\bm{)} P_\star(\bm{k}) \frac{d\bm{k}}{(2\pi)^2}
    \\
    & = \sum_{\star = \pm} \sum_{n, n', m} \int_{K_{n,n^\prime}^{(m)}}
    g\bm{(}\bm{v}_{\star}(\bm{k})\bm{)}P_{\star}(\bm{k}) \frac{d\bm{k}}{(2\pi)^2},
\end{align*}
where we have set 
\begin{equation*}
    P_{\star}(\bm{k}) := \left|\braket< \omega_{\star}(\bm{k}) |\hat{\Psi}_0(\bm{k}) >\right|^2, 
\end{equation*}
using the eigenstates 
$\ket |\omega_{\star}(\bm{k})>$ and 
group velocities $\bm{v}_{\star}(\bm{k})$
defined in Sec.~\ref{subsec:jointspectrum}.

Our strategy is to transform the $\bm{k}$-space integral on the final line of Eq.~\eqref{eq:spectral_integral_k} back into a 
$\bm{v}$-space integral over the joint spectrum $\Sigma(\hat{\bm{v}})$;
the resulting \emph{weight} function that multiplies
$g(\bm{v})$
must then be our PDF, $\rho(\bm{v})$.
This change of variable is precisely 
what our preparatory work in 
Secs.~\ref{sec:Jacodet}--\ref{sec:inversevelocitymap} was for.

To implement our strategy, we must express $P_{\star}({\bm k})$ as a function of $\bm{v}$ using the inverse velocity maps $\bm{k} = \bm{k}(\bm{v})$ constructed in Sec.~\ref{sec:inversevelocitymap}. This motivates us to introduce a simplified index 
$\Lambda = (\mu, \nu, \xi) \in \{0, 1\}^3$, whose components are defined to specify one of the eight covers from Theorem~\ref{thm:multicoveredjoint}:
\begin{itemize}
    \item[(1)] The index $\mu \in \{0, 1\}$ specifies the group of covers, corresponding to the LHS and RHS of the equations in Theorem~\ref{thm:multicoveredjoint}.
    We define the corresponding $n$-index set as
    $N_{\mu} = \{n + 4\mu \mid n = 1, 2, 3, 4\}$.
    \item[(2)] The indices $(\nu, \xi) \in \{0, 1\}^2$ specify the original index $m$, which determines the sign of the Jacobian (Proposition~\ref{prop:boundaryCm}). 
        As $\nu$ is an auxiliary index, 
        we align the parity of $m$ with $\xi$ so that
        $J_{(-1)^m}^{-1} = J_{(-1)^{\xi}}^{-1}$. 
        This mapping is given by
        \begin{equation} \label{eq:ms2s3}
            m = 2(\nu+1) - \xi.
        \end{equation}
\end{itemize}
Each index $\Lambda$ thus uniquely specifies a $k$-space domain $K_{\Lambda} = \bigcup_{n \in N_{\mu}} \bigcup_{n'} K_{n,n^\prime}^{(m)}$, where $m$ is from \eqref{eq:ms2s3} and the $n'$ union is over the non-empty domains in TABLE~\ref{tab:emptyset}. 
The inverse map associated with this domain is 
\begin{equation*}
    {\bm k}_{\Lambda}({\bm v}) := ({\bm v}|_{K_{\Lambda}})^{-1}({\bm v})
    \quad \text{on} \quad \Sigma(\hat{{\bm v}}) = {\bm v}(K_{\Lambda}).
\end{equation*}

Apply the change of variables $\bm{k} \mapsto \bm{v}$ 
using these inverse maps $\bm{k} = \bm{k}_{\Lambda}(\bm{v})$
and the corresponding Jacobian 
(Proposition \ref{prop:boundaryCm}), 
$d\bm{k} = |J|_{(-1)^{\xi}}^{-1} d\bm{v}$,
we can calculate
the $P_+(\bm{k})$ term in Eq.~\eqref{eq:spectral_integral_k}:
\begin{align}    
    & \sum_{\Lambda} \int_{K_{\Lambda}} g\bm{(}\bm{v}_+(\bm{k})\bm{)} P_+(\bm{k}) \frac{d\bm{k}}{(2\pi)^2} \notag \\
    & = \sum_{\Lambda} \int_{\Sigma(\hat{\bm{v}})} g(\bm{v}) P_+\bm{(}\bm{k}_{\Lambda}(\bm{v})\bm{)} |J|_{(-1)^{\xi}}^{-1} \frac{d\bm{v}}{(2\pi)^2} \notag \\
    & = \int_{\Sigma(\hat{\bm{v}})} g(\bm{v}) 
    \left[
        \sum_{\Lambda} P_+\bm{(}\bm{k}_{\Lambda}(\bm{v}) \bm{)}f_{(-1)^{\xi}}(\bm{v})
    \right] \dfrac{d\bm{v}}{4}.
    \label{eq:intP+term} 
\end{align}

To evaluate the $P_-(\bm{k})$ term, we must perform an additional change of variables $\bm{v} \mapsto -\bm{v}$. 
This transformation is facilitated by the symmetries of the velocity map and the ``windmill'' integration domains, 
as detailed below.

First, Eqs.~\eqref{defeq:v}--\eqref{vpm}
provide $\bm{v}_-(\bm{k}) = -\bm{v}(\bm{k})$.
Moreover, the $\bm{v} \mapsto -\bm{v}$ transformation in the velocity space
(which maps $g(-\bm{v}) \to g(\bm{v})$)
corresponds to a $\bm{k} \mapsto -\bm{k}$ transformation 
in the wave vector space.

Next, we see from the ``windmill'' structure in FIGs.~\ref{fig:only_K}--\ref{fig:windmill_waved_lines} that the $\bm{k} \mapsto -\bm{k}$ map corresponds to a $\pi$-rotation. This rotation permutes the integral domains $K_{n,n^\prime}^{(m)}$ according to the permutation ${\varepsilon} = (13)(24)(57)(68) \in S_8$ on the index $n$. Specifically, the index is transformed as $n \to n + 2$ for $n \in \{1, 2, 5, 6\}$ and $n \to n - 2$ for $n \in \{3, 4, 7, 8\}$, leaving the set of domains as a whole unchanged
Moreover, TABLE~\ref{tbl:sgn_s1_s2} and \eqref{defeq:v} show that
the permutation $\varepsilon$ makes the sign of $\bm{v}$ opposite;
the change of variable $\bm{v} \mapsto -\bm{v}$ leaves $P_-\bm{(}\bm{k}_{\Lambda}(\bm{v})\bm{)}$
invariant.

Applying these observations for calculating
the $P_-(\bm{k})$ terms in Eq.~\eqref{eq:spectral_integral_k}
yields:
\begin{align}
    & \sum_{\Lambda} \int_{K_{\Lambda}} g\boldsymbol{(}\bm{v}_-(\bm{k})\boldsymbol{)} P_-(\bm{k}) \frac{d\bm{k}}{(2\pi)^2} \notag \\
    = & \sum_{\Lambda} \int_{\Sigma(\hat{\bm{v}})} 
    g(-\bm{v}) P_-\bm{(}\bm{k}_{\Lambda}(\bm{v}) \bm{)}|J|_{(-1)^{\xi}}^{-1}(\bm{v}) \frac{d\bm{v}}{(2\pi)^2} \notag \\
    = &\int_{\Sigma(\hat{\bm{v}})} g(\bm{v}) 
    \left[
        \sum_{\Lambda} P_-\bm{(}\bm{k}_{\Lambda}(\bm{v}) \bm{)}f_{(-1)^{\xi}}(\bm{v})
    \right] \dfrac{d\bm{v}}{4}. \label{eq:intP-term} 
\end{align}

We now introduce functions  that incorporate
the results from Eqs.~\eqref{eq:intP+term}--\eqref{eq:intP-term}.
To streamline the notation, 
we relabel our eight inverse map 
$k_{\Lambda}$ with $\Lambda = (\mu, \nu, \xi)$ by
setting $\lambda = (\mu, \nu)$
and letting the index $\xi$
correspond to a new index $\bullet=\pm$.
We define $\bm{k}_{\lambda}^\bullet=\bm{k}_{\lambda}^\bullet(\bm{v})$ as
\begin{equation}
    \bm{k}_{\lambda}^{\bullet}
    = \begin{cases}
        \bm{k}_\lambda^{+}, &  \xi = 0, \\
        \bm{k}_\lambda^{-}, &  \xi = 1.
    \end{cases}
\label{eq:m3bullet}
\end{equation}
With this notation,
for $\star, \bullet \in \{+,-\}$, 
we define the four weight functions $w_{\star\bullet}$
as the sum over the first two indices:
\begin{align*}
    w_{\star\bullet}(\bm{v})
    = \sum_{{\lambda} \in \{0,1\}^2} 
    P_\star\bm{(}\bm{k}_{\lambda}^\bullet(\bm{v})\bm{)}.
\end{align*}

Adding Eqs.~\eqref{eq:intP+term}--\eqref{eq:intP-term}
and using the above notation
provides the following:
\begin{align}
    \eqref{eq:spectral_integral_k}
    & =\int_{\Sigma(\hat{\bm{v}})}g(\bm{v})
    \sum_{\star=\pm} \sum_{\bullet=\pm} w_{\star\bullet}(\bm{v})f_{\bullet}(\bm{v})d\bm{v}
     \notag \\
    & = \int_{\Sigma(\hat{\bm{v}})}g(\bm{v})
   \sum_{\bullet=\pm} w_{\bullet}(\bm{v})f_{\bullet}(\bm{v})d\bm{v},
    \label{eq:mainresult}
\end{align}
where 
\begin{equation} \label{eq:w_bullet}
    w_\bullet(\bm{v}):= \sum_{\star = \pm}  w_{\star\bullet}(\bm{v})
    = \sum_{\star = \pm}\sum_{\lambda \in \{0, 1\}^2} 
    P_{\star}\bm{(}\bm{k}_{\lambda}^\bullet(\bm{v})\bm{)}.
\end{equation}
This concludes the proof of Theorem~\ref{thm:WLT}.

Finally, we address the case $(a, b) \in \Delta_1$. 
As established in Proposition~\ref{prop:boundaryCm} (2), 
this regime is degenerate:
the domains $C^{(1)}$ and $C^{(3)}$ collapse onto the line $c_2 = -c_1$
(see FIG.~\ref{fig:C_m}),
whose contribution to the integral vanishes, 
leaving only the domains $C^{(2)}$ and $C^{(4)}$. 
According to our indexing \eqref{eq:ms2s3}, 
these domains $C^{(m)}$ ($m = 2, 4$) correspond to the index $\xi = 0$. 
From the definitions in \eqref{eq:m3bullet}--\eqref{eq:w_bullet}, $\xi = 0$ selects only the $\bullet = +$ branch.
Therefore, the $f_-$ and $w_-$ contributions vanish entirely.
The limiting PDF in Eq.~\eqref{eq:mainresult}
thus simplifies to $\rho(\bm{v}) = w_+(\bm{v})f_+(\bm{v})$.
Lemma \ref{lem:EFaresame} shows $f_+(\bm{v}) = f(\bm{v})$ for $v_{\mathrm{max}}=1$, 
which completes the proof of Theorem \ref{thm:WLT'}.

TABLE~\ref{tbl:difference} summarizes the differences between $v_{\mathrm{max}} < 1$ and $v_{\mathrm{max}} = 1$.
\begin{table}[ht]
	\caption{Difference between $v_{\mathrm{max}}<1$ and $v_{\mathrm{max}} = 1$} \label{tbl:difference}
	\begin{ruledtabular}
		\begin{tabular}{c|cc}
			   & $v_{\mathrm{max}}<1$ & $v_{\mathrm{max}}=1$ \\ \hline
			$\tau$ & $\max|\tau|<1$ & $\max|\tau|=1$ \\
            $\dfrac{\partial}{\partial k_j}\ket |\omega_\star(\bm{k})>$ & bounded & unbounded \\
			$v_j(\bm{k})$ & differentiable & differentiable without 4 points \\
			$j_{\pm}$ & $j_- < -1 < j_+$ & $j_- = j_+ = -1$ \\
            $F_j$ & $F_1 \neq F_2$ & $F_1 = F_2 = F_0$ \\
		\end{tabular}
	\end{ruledtabular}
\end{table}

We add one technical remark. Our proof derives the form of the limiting PDF, assuming the existence of the WLT. 
A standard sufficient condition for the WLT
—the boundedness of the eigenstate derivatives—
is satisfied for $v_{\mathrm{max}} < 1$ ($\Delta_{(0, 1)}$) 
but fails for $v_{\mathrm{max}} = 1$ ($\Delta_{1}$). 
This failure occurs because the derivatives diverge
at the four points where $|\tau(\bm{k})| = 1$. 
Nevertheless, the WLT itself is known to hold even in this degenerate case \cite[Technical remark at the end of Sec.~III]{grimmettWeakLimitsQuantum2004}. 
Our derivation of the PDF via integral transformation is therefore justified for all regimes discussed.

\section{CONCLUSION} \label{sec:conclusion}

In this work, we have derived the first exact representation of the limiting PDF for the WLT of a general 2D2SQW by introducing the notion of \emph{maximal speed} 
($v_{\rm max}$) to classify the dynamics.

Our central achievement is the discovery of \emph{2D Konno functions}, $f_{\pm}$ for the previously unexplored regime $\Delta_{(0, 1)}$ (i.e., $v_{\mathrm{max}} \in (0, 1)$),
which we establish as the \emph{proper} 2D generalization
of the 1D Konno function $f_{\mathrm{K}}$ 
by demonstrating their convergence
to $f_{\mathrm{K}}$ in the appropriate 1D limit. 
Furthermore, our formulation naturally recovers the previously known function $f$ for the $\Delta_1$ regime 
where $v_{\mathrm{max}} = 1$ as a degenerate case.

We also derive the closed-form expression for \emph{weight functions} $w_{\pm}$; together these provide a complete description of the limit distribution.

These results resolve a long-standing challenge in the study of QWs:
the explicit generalization of the Konno distribution to higher dimensions, a task that had remained elusive for general models. 

Methodologically, our success depends on completely resolving the non-injective velocity map. 
By partitioning the wave vector space based on
the zeros of the Jacobian,
which define the \emph{caustics},
we overcame the complexities of singular asymptotics inherent to 2D models.

Our approach provides a foundation for exploring extensions
to other 2D models, such as QWs on different lattice structures or those with multi-state coins. 
While extension to three or more dimensions presents significant analytical challenges, 
our work clarifies the necessary spectral analysis.

Finally, the relationship between the asymptotic behavior of our discrete (spacetime) QW model and the dynamics of a continuum analog, such as the massive Dirac equation 
in $(2 + 1)$-dimensions, remains an intriguing direction.
Our 2D Konno functions ($v_{\mathrm{max}} < 1$) may serve as a signature of the mass term, contrasting with the massless dynamics of the $v_{\mathrm{max}}=1$ regime.

\begin{acknowledgments}
We thank C.~Cedzich for his helpful remarks and for bringing references \cite{ahlbrechtAsymptoticEvolutionQuantum2011, cedzichExponentialTailEstimates2024} to our attention.
This work was supported by JSPS KAKENHI Grant Numbers JP23K03229 and	JP24K06851 and by the 
Research Institute for Mathematical Sciences,
an International Joint Usage/Research Center located in Kyoto University.
\end{acknowledgments}

\appendix

\section{Proofs of Lemmas}
\subsection{Proof of Lemma \ref{lem:equiv}}
\label{appen:pr:lem:equiv}
 The relations $e^{-i \bm{\theta} \cdot \bm{x}} S_j e^{i \bm{\theta} \cdot \bm{x}} 
    = e^{i \theta_j \sigma_z} S_j
    = S_j e^{i \theta_j \sigma_z}$,
and
$
C_j^\prime 
= e^{i\delta_j/2} e^{i\psi_j \sigma_z} C_je^{i\varphi_j \sigma_z}
$
 with $\psi_j = (\alpha_j+\beta_j)/2$ and $\varphi_j = (\alpha_j-\beta_j)/2$
 prove
\begin{align*}
    e^{-i(\delta_1 + \delta_2)/2} U^\prime
   = \left(e^{-i\varphi_1 \sigma_z} e^{i \theta \cdot x} \right)
    U 
    \left(e^{-i \theta \cdot x} e^{i\varphi_1 \sigma_z}\right)
\end{align*}
with $\bm{\theta} = (\varphi_1+\psi_2, \psi_1+\varphi_2)$. 
Taking $\gamma=(\delta_1+\delta_2)/2$ and
$W = e^{-i\varphi_1 \sigma_z} e^{i \theta \cdot x} $,
we obtain (i). 

Using the spectral decomposition $\hat{\bm{x}} = \sum_{\bm{x} \in \mathbb{Z}^2} \bm{x} \hat{\Pi}_{\bm{x}}$
of the position operator, one observes that
$\Psi_t^\dagger(\bm{x}) \Psi_t (\bm{x}) = \langle \Psi_t, \hat{\Pi}_{\bm{x}} \Psi_t \rangle_{\mathcal{H}}$.
Combining this with (i) yields (ii)
because $W$ commutes with $\hat{\Pi}_{\bm{x}}$,
i.e., $[W, \hat{\Pi}_{\bm{x}}] = 0$. 

\subsection{Proof of Lemma \ref{lem:a1a2toab}}
\label{appen:pr:lem:a1a2toab}
    We only prove that $\Phi_{12}$ is a one-to-one onto correspondence.
    Because $(a, b) \in \Delta$ satisfies 
    $0 \leq a + b \leq 1$, $-1 \leq a - b \leq 1$ and $|a - b|\leq a + b$,
    one can define
    \begin{align}
    \label{def:theta1}
    &\theta_1 = \frac{1}{2}\bm{(}\arccos(a - b) + \arccos(a + b)\bm{)}, \\
    \label{def:theta2}
    &\theta_2 = \frac{1}{2}\bm{(}\arccos(a - b) - \arccos(a + b)\bm{)},
    \end{align}
    so that  $\theta_j \in [0, \pi/2]$ ($j = 1, 2$).  
    Then $a_j:= \cos \theta_j$ and $b_j:= \sin \theta_j$
    satisfy $a = a_1a_2$, $b = b_1b_2$, and $(a_1, a_2) \in A_{12}$. 
    Thus $\Phi_{12}$ is an onto map. 
    
    Suppose that there exists another pair $(a_1^\prime, a_2^\prime) \in A_{12}$
    so that $a = a_1^\prime a_2^\prime$.  
    Then $\theta_j^\prime = \arccos a_j^\prime \in [0,\pi/2]$ ($j = 1, 2$) 
    satisfy $b_j^\prime = \sin \theta_j^\prime$ ($j = 1, 2$) and
$        a \pm b = \cos (\theta_1^\prime \mp \theta_2^\prime).
$
    Substituting these equations 
    into the RHS of \eqref{def:theta1} and \eqref{def:theta2} yields $\theta_j^\prime = \theta_j$ ($j = 1, 2$). 
    Thus $\Phi_{12}$ is a one-to-one map. 
    Therefore we obtain the lemma. 

\subsection{Proof of Lemma \ref{lem:5}}
\label{app:pr:lem:5}
    By assumption, $v_{\mathrm{max}} = a+b < 1$. 
    (1) follows by definition. 
    By the arithmetic and geometric mean relation,
    \begin{align*} 
    F_0-\sqrt{F_1F_2}
        & \geq F_0 - (F_1+F_2)/2 \\
        & \geq [1- (a+b)][1-(a-b)^2]/(2ab)
    \end{align*}
    holds in $E_1 \cap E_2$. 
    (2) follows since the LHS of the above inequality
    is strictly positive
    by assumption.
    The facts that $E_1 \cap E_2 \subset \sigma(\hat{v}_1) \times \sigma(\hat{v}_2
    )$
    and hence that 
    \begin{equation*}
    (1 - v_1^2)(1 - v_2^2) > (1 - v_{\mathrm{max}}^2)^2 
    \end{equation*} 
    in $E_1 \cap E_2$ prove (3). 
\section{Proof of Proposition \ref{prop:f_Kemerges}}
\label{app:pr:prop:f_Kmetges}

Let $g = \tilde{h} \tilde{K}_{\pm}$ and
$\displaystyle I_j =\iint_{\tilde{D}_j} g\, dvdu$
for $j = 1, 2$.

We first prove $\lim_{b \to 0}I_1 = 0$. 
Lemma \ref{lem:5} states that
$K_{\pm} \leq F_0 / \sqrt{F_1F_2}$ and $ab F_0 \leq 1$ on $E_1 \cap E_2$.
Using the identity $(a_jb_j)^2 \tilde{F}_j 
= a_j^2(r_j^2 - u^2)$ ($j = 1, 2$) obtained by definition
yields the following inequality:
\[
    g 
    \lesssim 
    \dfrac{1}{ab\sqrt{\tilde F_1 \tilde F_2}}
    = \dfrac{1}{a\sqrt{(r_1^2 - u^2)(r_2^2 - u^2)}}
\]
on $\tilde{D}_1 \cup \tilde{D}_2$,
where $g_1 \lesssim g_2$ for functions $g_1$ and $g_2$ indicates 
that $g_1/g_2$ is bounded above by a constant
independent of the variables of $g_1$ and $g_2$. 
Observe that
\[ 
    r_2^2-u^2
    \geq r_2^2- r_1^2 =\frac{b_1^2- b_2^2}{a^2} (v^2 -a^2) > 0
    \quad \mbox{ on $\tilde{D}_1$}. 
\]
Combining these inequalities 
and using 
the symmetry of $\tilde{D}_j$ and the integrand
yield:
\begin{align*}
    I_1 \lesssim 
    \int_a^{a_1} \dfrac{dv}{\sqrt{v^2-a^2}} 
    \int_0^{r_1}\dfrac{du}{\sqrt{r_1^2-u^2}} 
    =
    \frac{\pi}{2} 
    \log \left(1 + \dfrac{b_2}{a_2} 
    \right), 
\end{align*} 
where the RHS tends to zero as $b_2 \to 0$.
The last equality follows
from the following formulae (up to constants) 
\[ 
    \int \dfrac{dx}{\sqrt{|x^2 - s^2|}}
    =\begin{cases}
        \arcsin \dfrac{x}{s}, & s > |x|, \\
        \log \left(x + \sqrt{x^2 - s^2}\right) , & |x| > |s|.
    \end{cases}
\]

We next examine the limit of $I_2$ as $b \to 0$. 
Observe that $r_2$ tends to zero as $b_2 \to 0$
and that $b_2^2\tilde{F}_2(v, r_2u_2) = r_2^2(1 - u_2^2)$. 
The change of variables $u_2 = u/r_2$
yields
\begin{align*}
    I_2 & = r_2 \int_{-a}^{\ a} dv \int_{-1}^1 du_2 \, \tilde{h}(v,r_2 u_2) \tilde{K}_{\pm}(v,r_2 u_2) \\
    & \sim \frac{b_1}{4} \int_{-a_1}^{\ a_1} dv \, \frac{(1 - v^2) \tilde{h}(v, 0)}{\sqrt{a_1^2 -v^2}} 
    \times \int_{-1}^1 
    \frac{du_2}{\sqrt{1-u_2^2}}
\end{align*}
as $b \to 0$, where $A \sim B$ for constants $A$ and $B$ depending on $b$ 
denotes $|A-B|=o(b)$ as $b \to 0$. 
The desired result then follows from the Lebesgue convergence theorem and calculating the integral directly. 

\section{\texorpdfstring{Proof of Proposition \ref{prop:boundaryCm}}{Proof of Proposition 19}}
\label{pr:prop:boundaryCm}
In this proof,
the quadratic form $\varphi({\bm c})$ defined in \eqref{def:varphi}
plays a key role. We now define a single-variable quadratic polynomial $\varphi_1(x)$
by $\varphi_1(x) = \varphi(1, x)$.
The discriminant $D_1= [1 - (a - b)^2][1 - (a + b)^2]$ of $\varphi_1(x)$ 
is always non-negative if $a + b \leq 1$;
$D_1$ attains zero only when $a+b=1$. 
Thus, $\varphi_1$ has two zeros if $a + b < 1$
and one zero if $a + b = 1$;
Writing the zeros as $j_{\pm}$ with ordering $j_+ \geq j_-$
observes that the lines $c_2 = j_\pm c_1$ 
partition the square $[-1, 1]^2$ in the $c_1c_2$-plane 
into four regions (as depicted in FIG.~\ref{fig:C_m})
$C^{(m)}$ ($m=1, 2, 3, 4$) if $a + b < 1$,
and into two regions $C^{(2)}, C^{(4)}$ if $a + b = 1$.
The factorization $\varphi_1(x) = ab(x - j_+)(x - j_-)$ 
and the identity $\varphi({\bm c}) = c_1^2 \varphi_1(c_2/c_1)$
yield the identity
\begin{equation*}
    \varphi({\bm c}) 
    = ab \left(c_2 - j_+ c_1 \right) 
    \left(c_2 - j_- c_1 \right),
\end{equation*}
which determines the sign of the quadratic form $\varphi({\bm c})$ and proves that $\varphi({\bm c}) = 0$ holds only 
on the lines $c_2 = j_\pm c_1$;
otherwise $\sgn \varphi({\bm c}) = (-1)^m$ on $C^{(m)}$.
Combining this with the relation \eqref{repr:Jacobdet}
ensures that $J = 0$ on the lines $c_2 = j_\pm c_1$,
otherwise $|J|^{-1}=|J|^{-1}_{(-1)^m}$

\section{\texorpdfstring{Proof of Theorem \ref{thm:jointspectrum} 
}{Idea of the proof of Sigma(hat v) = E1 cap E2}}\label{appen:sigma_v}
In this section, we determine the joint spectrum 
$\Sigma(\hat{\bm v})$, which is the closure of the image of the velocity map ${\bm v}({\bm k})$. 
We focus on the $(a,b)\in\Delta_{(0,1)}$ case (where $v_{\mathrm{max}} < 1$); 
the $\Delta_1$ case
(where $v_{\mathrm{max}} = 1$)
is addressed at the end.

\begin{figure}
	\includegraphics[width=5cm]{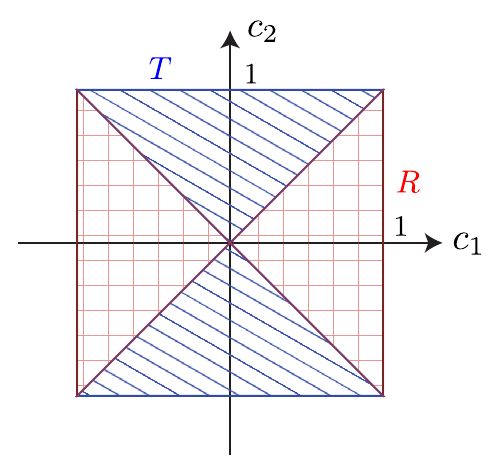}
	\caption{Division of $[-1, 1]^2$ into $R$ and $T$ in $c_1c_2$-plane. The region shaded with red grids in the shape of a ribbon represents $R$, while the region shaded with blue diagonal lines in the shape of a twister represents $T$.} \label{fig:Ribbon_Twist}
\end{figure}

The proof becomes clear with the change of variables 
$\bm{u}= (v_1+v_2, v_1-v_2)/\sqrt{2}$, 
which corresponds to a standard rotation of the coordinate system
by $\pi/4$. 
Eq.~\eqref{defeq:v} then provides
\begin{equation}
u_1 = 
    -\dfrac{\sqrt{2}as_1}{\sqrt{1 - \tau^2}}, 
\quad u_2 = 
    -\dfrac{\sqrt{2}bs_2}{\sqrt{1 - \tau^2}}.   
    \label{defeq:u}
\end{equation}
Here, $\bm{u}$ depends on ${\bm c}=(c_1, c_2)$
and the signs of $s_i$. 
Since the signs of $s_i$ (and thus the signs of $u_1, u_2$) are uniquely determined by the index $n$ via Table~\ref{tbl:sgn_s1_s2}, 
the map $\bm{u}$ exhibits symmetry across the quadrants.
To analyze the image, we follow the partitioning scheme introduced in Sec.~\ref{subsubsec:Knm}. We first partition the $[-1, 1]^2$ in the $c_1c_2$-plane into the regions $R$ and $T$ (see FIG.~\ref{fig:Ribbon_Twist}):
\begin{align*}
    R &:= \big\{ {\bm c} = (c_1, \kappa c_1) \mid 
        c_1 \in [-1,1], \kappa \in [-1, 1] \big\}, \\
    T &:= \big\{ {\bm c} = (\gamma c_2, c_2) \mid 
        c_2\in [-1,1], \gamma \in [-1, 1] \big\}.
\end{align*}

\begin{figure}
	\includegraphics[width=8cm]{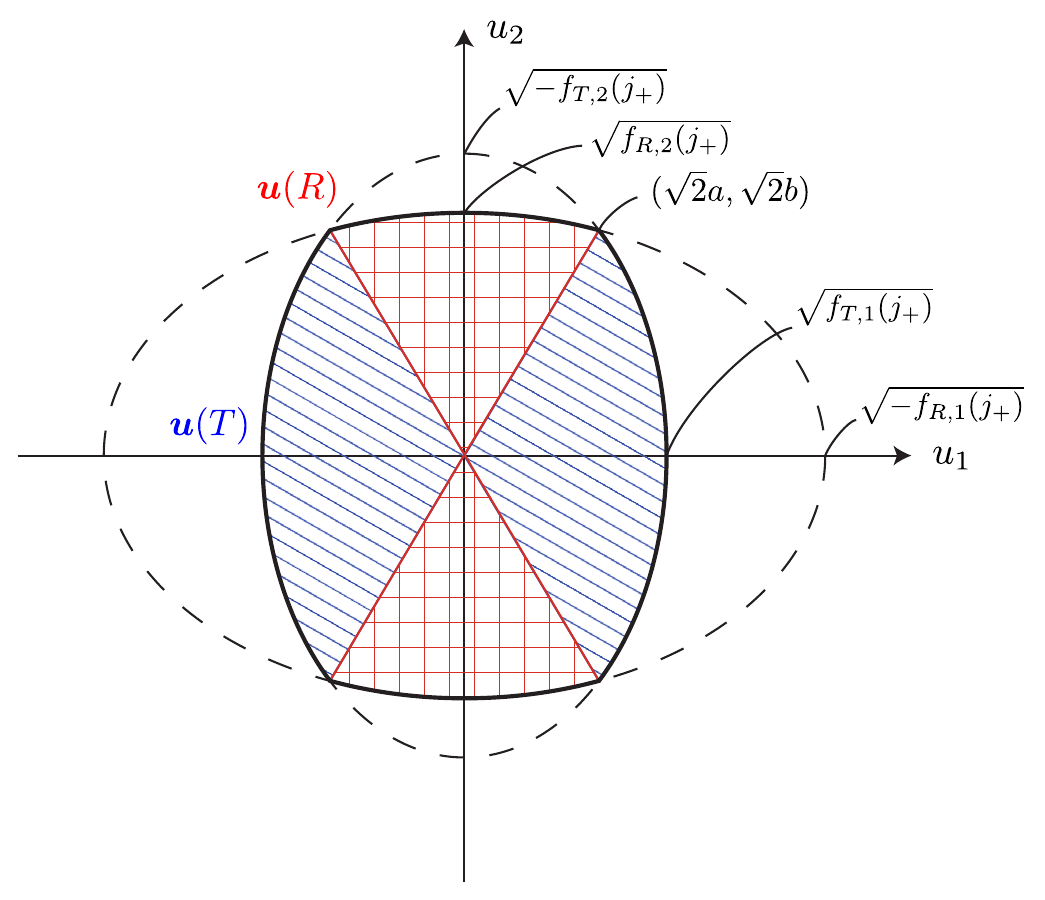}
	\caption{
     The joint spectrum $\Sigma(\hat{\bm v})$ visualized in the $\pi/4$-rotated coordinate system ${\bm u}=(u_1, u_2)$.
     The full domain (shaded) is the intersection of two ellipses, corresponding to $E_1 \cap E_2$ in FIG.~\ref{fig:E1_cap_E2}. It is partitioned into the image ${\bm u}(R)$ (red grids) and ${\bm u}(T)$ (blue diagonal lines), which originate from the $R$ and $T$ regions defined in FIG.~\ref{fig:Ribbon_Twist}. The dashed curves show the boundaries of the two individual ellipses ($E_1$ and $E_2$) . These boundaries are generated by the critical values $\kappa=j_{+}$ or $\gamma=j_{+}$, where $\kappa\in[-1,1]$ is a parameter of \eqref{eq:curve_u} and so is $\gamma$, and their maximal extents are marked by the labeled axis intercepts (e.g., $\sqrt{f_{T,1}(j_+)}$ and $\sqrt{f_{R,2}(j_+)}$).
    } \label{fig:sigma_v_hat}
\end{figure}

The domains $K_{n,s}$ (for $s=R, T$) are then defined as the pullbacks of these regions into the $n$-th windmill domain $K_n$.
The labels $\bm{u}(R)$ and $\bm{u}(T)$ in FIG.~\ref{fig:sigma_v_hat} correspond to the union of the images: 
\[ \bm{u}(R) = \bigcup_{n=1}^4 \bm{u}(K_{n,R}),
    \quad \bm{u}(T) = \bigcup_{n=1}^4 \bm{u}(K_{n,T}). \]

\begin{figure*}
	\includegraphics[width=17.8cm]{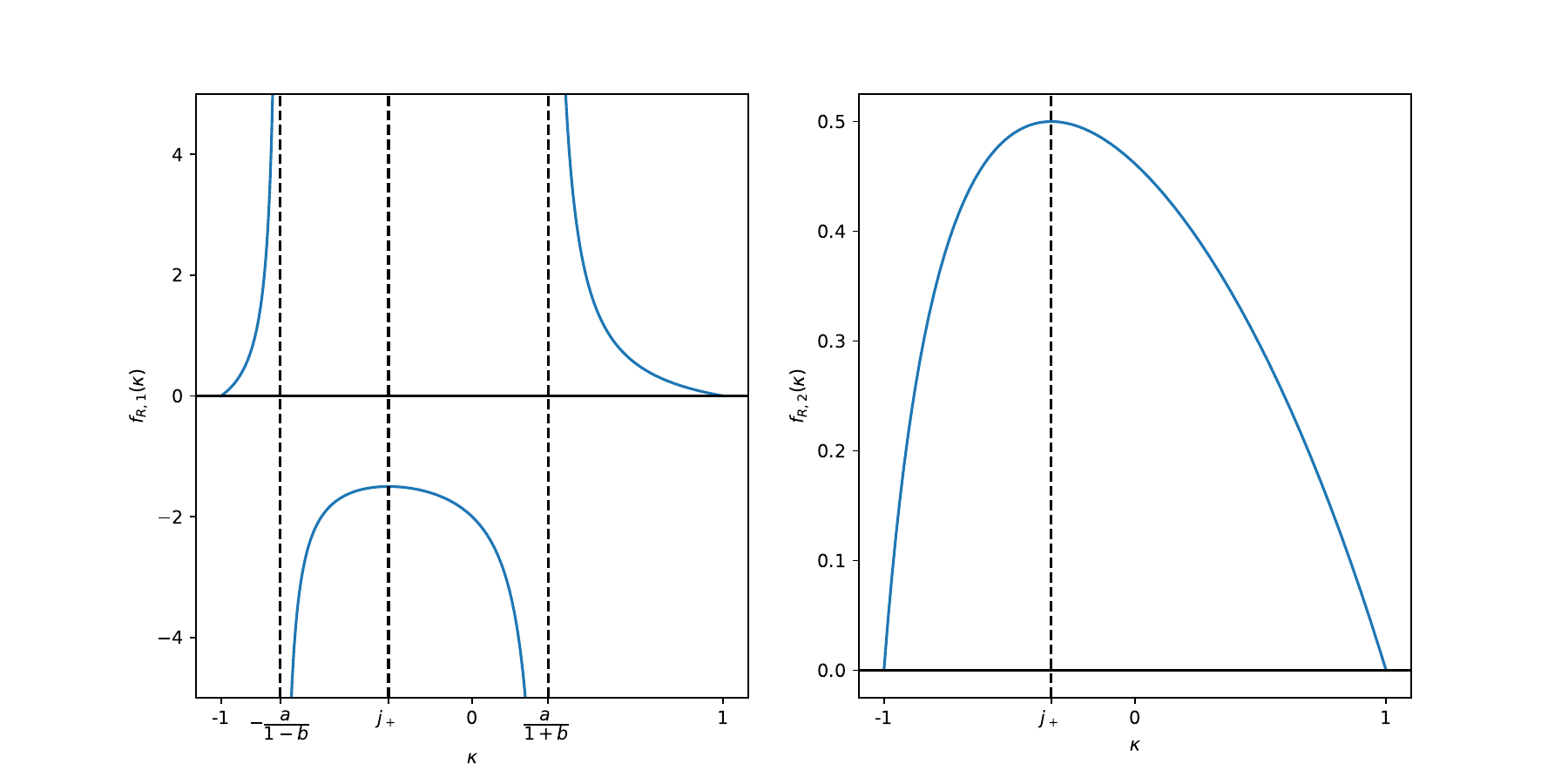}
	\caption{
    Graphs of the functions $f_{R,1}(\kappa)$ (left) and $f_{R,2}(\kappa)$ (right), defined as
    $f_{R, 1}(\kappa) 
    = [2a^2(1 - \kappa^2)]/[\kappa^2 - (a - b\kappa)^2]$, 
    $f_{R, 2}(\kappa) 
    = [2b^2(1 - \kappa^2)]/[1 - (a - b\kappa)^2]$ for $\kappa \in [-1, 1]$.
    These functions determine the geometry of the image ${\bm u}(R)$ via \eqref{eq:curve_u}. 
    The plot of $f_{R,1}$ illustrates its alternating sign, diverging at the vertical asymptotes $\kappa = -a/(1-b)$ and $\kappa = a/(1 + b)$. In contrast, $f_{R,2}(\kappa)$ is strictly positive for $\kappa \in (-1, 1)$.
    Both functions have a maximal at $\kappa = j_+$—specifically, a maximum for $f_{R, 2}(\kappa)$ and a local maximum for $f_{R, 1}(\kappa)$—which corresponds to the maximal extent of the image $\bm{u}(R)$.
    } \label{fig:f_Ribbon}
\end{figure*}

We will now show that the union of the images $\bm{u}(R)$ and $\bm{u}(T)$ corresponds to the $\pi/4$-rotation of $E_1 \cap E_2$, as illustrated in FIG.~\ref{fig:sigma_v_hat}. 
To analyze the image ${\bm u}(R)$,
we square two equations in \eqref{defeq:u}
and substitute $c_2=\kappa c_1$ into them. 
Eliminating $c_1^2$ from these equations yields:
\begin{equation}
    \frac{u_1^2}{-f_{R,1}(\kappa)}+\frac{u_2^2}{f_{R,2}(\kappa)}=1, \label{eq:curve_u}
\end{equation}
where $f_{R,1}$ and  $f_{R,2}$ are functions of the parameter $\kappa$.
The explicit definitions of these functions and their analysis are provided in FIG.~\ref{fig:f_Ribbon}. 
As shown there, \eqref{eq:curve_u} describes ellipses (when $f_{R,1}<0$) or hyperbolas (when $f_{R,1}>0$). As $\kappa$ varies, the union of these curves fills the regions $\bm{u}(R)$. 
The values $\kappa=\pm1$ define the boundaries between ${\bm u}(R)$ and ${\bm u}(T)$, while $\kappa=j_{+}$ 
(given Sec.~\ref{subsubsec:Knm}) 
provides the maximal extent of the full image.

These functions $f_{R,1}$ and $f_{R,2}$, at this critical point, take
$f_{R,1}(j_+) = -2(a_1 \!\vee\! a_2)^2$ and $f_{R,2}(j_+) = 2(b_1 \wedge b_2)^2$,
and hence \eqref{eq:curve_u} forms the boundary ellipse:
\begin{equation}
    \frac{u_1^2}{2(a_1 \!\vee\! a_2)^2}+\frac{u_2^2}{2(b_1\wedge b_2)^2}=1,
    \label{eqC5}
\end{equation}
where $x \wedge y := \min\{x, y\}$ and $x \vee y := \max\{x, y\}$.
By similar calculations for $T$, the boundary $\gamma=j_{+}$ (i.e., $c_2=j_- c_1$) yields the other ellipse:
\begin{equation}
    \frac{u_1^2}{2(a_1\!\wedge\!a_2)^2} + \frac{u_2^2}{2(b_1\!\vee\!b_2)^2} = 1.
    \label{eqC6}
\end{equation}
Since $\bm{u}(R) \cup \bm{u}(T)$ is
the intersection of two ellipses bounded by \eqref{eqC5} and \eqref{eqC6},
it is precisely the $\pi/4$-rotated image of $E_1 \cap E_2$, 
which proves $\Sigma(\hat{\bm v}) = E_1 \cap E_2$.

When $(a,b) \in \Delta_1$, 
the parameters degenerate as $j_\pm=-1$. 
This is a crucial difference from the $(a,b) \in \Delta_{(0,1)}$ case: 
the caustic boundary (i.e., $\kappa=j_+$) 
now merges with the $R/T$ partition boundary $\kappa = -1$.
In this degenerate case the caustic boundary no longer maps to a curve, but collapses to the four discrete points 
$\pm(\sqrt{2}a, \sqrt{2}b)$ and $\pm(\sqrt{2}a, -\sqrt{2}b)$, which 
are located at the boundary $u_1^2/(2a) + u_2^2/(2b)=1$
of the $\pi/4$-rotated image of the ellipse $E$.
The other boundary $\kappa = 1$ and the entire interior of
$[-1,1]^2$ in the $c_1c_2$-plane both map to the open interior of the above ellipse. 
Therefore, the image of the map ${\bm u}={\bm u}({\bm c})$
is the open single elliptical region plus four discrete boundary points. As the joint spectrum 
must be closed, we observe $\Sigma(\hat{\bm v})=E$.

\section{Derivation of 1D Limits}
\label{sec:1Dlimits}
Here we sketch the proofs of Theorems~\ref{thm:btozero} and~\ref{thm:btozero1}.
The proofs are based on the fact that
the 2D velocity operators $\hat v_1$ and $\hat v_2$
both converge to the direct sum of copies of the 1D velocity operator 
$\hat v_{\mathrm{K}}$ in the 1D limit $b \to 0$.

Recall that the 1D velocity operator $\hat v_{\mathrm{K}}(k)$
for a translation invariant 1D2SQW with the evolution 
$\hat U_{\mathrm{K}}(k) 
= \begin{pmatrix}
e^{ik} & 0 \\  0 & e^{-ik}    
\end{pmatrix}  \begin{pmatrix}
a_1 & b_1 \\  -b_1 & a_1    
\end{pmatrix}$ in the wave number space $\mathbb{T}$
is defined as follows \cite{grimmettWeakLimitsQuantum2004, suzukiAsymptoticVelocityPositiondependent2016}.
Let $e^{\pm i \omega_{\mathrm{K}}(k)}$ be
the eigenvalue of $\hat U_{\mathrm{K}}(k)$, 
where the dispersion relation is
$\omega_{\mathrm{K}}(k) = \arccos (a_1 \cos k)$. 
The 1D group velocity is then given by 
\begin{equation}
\label{defeq:vK}
   v_{\mathrm{K}}(k) 
    := -\dfrac{a_1 \sin k}{\sqrt{1-(a_1 \cos k)^2}},
\end{equation} 
which is consistent with the sign convention
of the group velocity as defined in Sec.~\ref{subsec:jointspectrum}.
The 1D velocity operator 
$\hat v_{\mathrm{K}}(k)$ is defined as
\begin{equation}
\label{defeq:hatvK}
\hat v_{\mathrm{K}}(k) 
= \sum_{\star=\pm} v_{\mathrm{K},\star}(k)  
    \ketbra | \omega_{\mathrm{K},\star}(k) > < \omega_{\mathrm{K},\star}(k) |, 
\end{equation} 
where $v_{\mathrm{K},\pm}(k) := \pm v_{\mathrm{K}}(k)$,
and the corresponding eigenvectors are
\begin{equation}
\label{eq:omegaK}
      \ket |\omega_{\mathrm{K},\pm}(k) > = \dfrac{1}{\sqrt{2}}
\begin{pmatrix} 
    e^{i k} \sqrt{1 \mp v_{\mathrm K}(k)} \\
    \pm i \sqrt{1 \pm v_{\mathrm K}(k)}
\end{pmatrix},
\end{equation}
which form an ONB of $\mathbb{C}^2$.

As explained immediately after Definition~\ref{def:1DL(01)},
the 2D evolution $U_{(a,b)}$ converges 
in the 1D limit to the 1D evolution $U_{a_1}$,
which is identical to the operator 
in the $\Delta_{ab=0}$ regime (see TABLE~\ref{tab:3}).  
As mentioned in Sec.~\ref{sec:reducibleQWs},
such an operator can decouple 
into an infinite direct sum of 1D motions,
each governed by the evolution $\tilde{S}_{\mathrm{1D}} C_1$.
Identifying 
$\ell^2(\mathbb{L}_n^{(+)};\mathbb{C}^2)$ 
with $\ell^2(\mathbb{Z};\mathbb{C}^2)$ 
shows that $\tilde{S}_{\mathrm{1D}} C_1$
is unitarily equivalent to $U_{\mathrm{K}}$,
i.e., the position representation of $\hat U_{\mathrm{K}}$
stated above. 
Implement this identification 
by $\psi_0^{(n)}(m) := \Psi_0(m,n+m)$
provides
\begin{equation}
\label{eq:1Ddecouple}   
\hat{\Psi}_0({\bm k}) 
= \sum_{n \in \mathbb{Z}} 
e^{-in k_2}\hat{\psi}_0^{(n)}(k_1+k_2),
\end{equation} 
where $\Psi_0 \in \mathcal{H}$ is the initial state
and $\hat{\psi}_0^{(n)}$ denotes the 1D Fourier transform of $\psi_0^{(n)}$.

The remainder of this section
uses the terminology defined in Eqs.~\eqref{defeq:l}--\eqref{vpm}, and \eqref{defeq:taupm}. 
To emphasize the dependence on $b$, 
we denote 
the 2D velocity operators 
by $\hat{\bm v}_b= (\hat v_{1,b}, \hat v_{2,b})$.
Similarly, we use ${\bm v}_b({\bm k}) = \bm{(}v_{1,b}({\bm k}),v_{2,b}({\bm k})\bm{)}$ to denote 
the 2D velocity map.

\subsection{1D Limits for Velocity Operators}
In this subsection,
we show that
the 1D limit $b \to 0$
decouples the 2D velocity operators
$\hat v_j$
into an infinite direct sum of 
copies of the 1D velocity operator $\hat v_{\mathrm{K}}$.
Specifically, we prove the following:
Let
\begin{equation}
\label{defeq:Pstarkn}
 P_{{\mathrm{K}},\star}^{(n)}(k)
 = \left| \braket< \omega_{\mathrm{K},\star}(k) 
    |\hat{\psi}_0^{(n)}(k) > \right|^2.
\end{equation}
\begin{prop}
\label{prop:decompgrho}
For any continuous function $g$,
\begin{align}
    & \lim_{b \to 0}
    \int_{\Sigma(\hat{\bm{v}}_b)} g(\bm{v})\, 
    d \braket < \Psi_0 | E_{\hat{\bm{v}}_b}(\bm{v}) \Psi_0> \notag \\
    & = \sum_{n \in \mathbb{Z}} \sum_{\star = \pm}
    \int_{-\pi}^\pi \dfrac{dk}{2\pi}
     g\bm{(}v_{{\mathrm{K}},\star}(k),v_{{\mathrm{K}},\star}(k)\bm{)}
     P_{{\mathrm{K}},\star}^{(n)}(k).
     \label{eq:1Dlimgvb}
\end{align}
\end{prop}
\begin{proof}
By definitions, the 1D limit $b \to 0$ (which implies $b=0$ and $a=a_1$) of the 2D velocity map
${\bm v}_b$ 
is a one-variable function of $l_1=k_1+k_2$:
$\lim_{b \to 0} v_{j,b}({\bm k})
= v_{\mathrm{K}}(l_1)$
with $v_{\mathrm{K}}$ defined in \eqref{defeq:vK}.
The eigenvectors of $\hat{v}_{j,b}$
are represented as 
\begin{equation*}
\ket |\omega_\pm ({\bm k})> 
= \dfrac{1}{\sqrt{2}}
\begin{pmatrix} 
    e^{i \eta_b({\bm k})}\sqrt{1 \mp v_{2,b}({\bm k})} \\
    \pm i \sqrt{1 \pm v_{2,b}({\bm k})}
\end{pmatrix}
\end{equation*}
with the argument $\eta_b({\bm k})$ 
of $a_2  b_1 e^{i l_1} + b_2 a_1 e^{i l_2}$.
By $\lim_{b \to 0} e^{i \eta_b({\bm k})} = e^{i l_1}$, 
we obtain
$\lim_{b \to 0}\ket|\omega_{\star}({\bm k})> 
= \ket|\omega_{\mathrm{K},\star}(l_1)>$, 
which are also functions of $l_1$.
Applying these limits to
the decomposition \eqref{def:velocityop}
shows the pointwise convergence 
$\lim_{b \to 0} \hat v_{j,b}({\bm k})
    =  \hat v_{\mathrm K}(l_1), 
$
where $\hat v_{\mathrm K}$ is the 1D velocity operator defined in \eqref{defeq:hatvK}.
Using the spectral decomposition
and \eqref{eq:1Ddecouple},
we observe that 
\begin{align}
& \text{LHS of \eqref{eq:1Dlimgvb}} \notag \\
& = \sum_{\star=\pm} \sum_{n,n'} 
    \int_{\mathbb{T}^2} \dfrac{d{\bm k}}{(2\pi)^2}
    g\bm{(}v_{{\mathrm K},\star}(l_1),v_{{\mathrm K},\star}(l_1)\bm{)} e^{i(n'-n)k_2} 
\notag \\
& \qquad \times 
\braket< \hat{\psi}_0^{(n')}(l_1) 
    |\omega_{\mathrm{K},\star}(l_1) >
\braket< \omega_{\mathrm{K},\star}(l_1) 
    |\hat{\psi}_0^{(n)}(l_1) >. \notag
\end{align}
After changing variables to $(k,k')=(k_1+k_2,k_2)$, periodicity ensures that $k'$ only appears in the factor $e^{i(n-n')k'} / (2\pi)$. 
Integrating this factor simplifies the sum over $n'$ and yields \eqref{eq:1Dlimgvb}. 
\end{proof}

\subsection{1D Limit in \texorpdfstring{$\Delta_{(a,b)}$}{Delta(a,b)} Regime}
We prove Theorem~\ref{thm:btozero}. 
By the definition of $\rho_{(a,b)}$
and Proposition~\ref{prop:decompgrho},
we need only calculate the RHS of \eqref{eq:1Dlimgvb}.
Following \cite[Lemma 4.6]{richardQuantumWalksAnisotropic2018b},
we define the 1D inverse velocity map $k=k_{\star,m}(v)$,
which appeared in \eqref{wv}, as follows. 
We divide $\mathbb{T}$ into $\mathbb{T}_0=[-\pi/2, \pi/2]$
and $\mathbb{T}_1=[\pi/2, 3\pi/2]$,
where the restrictions $v_{\mathrm{K}, \star}|_{\mathbb{T}_m}$
($\star=\pm$, $m=0,1$)
of the velocity map $v = v_{{\mathrm{K}},\star}$ are invertible.
The resulting inverse functions are given by
\begin{equation}
\label{defeq:kpmmv}
k_{\pm,m}(v)= m \pi 
    + \arcsin \left( \dfrac{\mp (-1)^m b_1 v}{a_1\sqrt{1-v^2}}  \right) 
\end{equation}
for $v \in [-a_1,a_1]$, 
whose derivatives, as mentioned at the end of Sec.~\ref{subsec:WLT}, provide the Konno function
$f_{\mathrm{K}}$.

Changing variables by $k=k_{\pm,m}(v)$ 
allows us to represent
\begin{align*}
\lim_{b_2 \to 0} \int_{\Sigma(\hat{\bm v})} g({\bm v}) 
    \rho_{(a,b)}({\bm v})d{\bm v}
= \int_{-a_1}^{a_1}
        g(v,v) \rho_{a_1}(v) dv,
\end{align*}
where $\rho_{a_1}(v) = w(v;a_1) f_{\mathrm{K}}(v;a_1)$
given in \eqref{rhsG_0} is the 1D limit of $\rho_{(a,b)}(v)$ and the weight $w(v;a_1)$ is given by
\begin{equation}
    w(v;a_1)=\frac{1}{2}\sum_{\substack{n \in {\mathbb Z} \\ \star=\pm, \ m = 0, 1}} 
     P_{\mathrm{K},\star}^{(n)}\bm{(}k_{\star,m}(v)\bm{)},
    \label{eq:weight_bto0}
\end{equation}
where $P_{\mathrm{K},\star}^{(n)}(k)$ and $k_{\star,m}(v)$
are defined in \eqref{defeq:Pstarkn} and \eqref{defeq:kpmmv}.
\subsection{1D Limit in \texorpdfstring{$\Delta_1$}{Delta1} Regime}
We prove Theorem~\ref{thm:btozero1}. 
The 1D limit in the $\Delta_1$ regime corresponds to the 1D QW model with $a_1=1$. 
We therefore evaluate the RHS of \eqref{eq:1Dlimgvb} for this specific case. 
Setting $a_1=1$ in \eqref{defeq:vK} simplifies 
the velocity to
$v_{\mathrm{K}}(k) = - \sgn(\sin k)$.
Combining this with \eqref{eq:omegaK} 
allows us to take (up to phase factors)
$\ket |\omega_{\mathrm{K},\pm }>$ as
\begin{equation*}
\begin{cases}
\ket |\omega_{\mathrm{K},\pm }(k)> = \ket |\mp >
     \qquad  
        & \text{if} \quad v_{\mathrm{K}}(k) = 1, \\
\ket |\omega_{\mathrm{K},\pm }(k)> = \ket |\pm >
    \qquad  
        & \text{if} \quad v_{\mathrm{K}}(k) = -1,
\end{cases}
\end{equation*}
where we define $\ket | \pm >$ so that $\braket  <\pm |\Psi(x) > = \Psi^{(\pm)}(x)$. 
Substituting this into \eqref{defeq:Pstarkn}
and calculating the RHS of \eqref{eq:1Dlimgvb},
we use the fact that $v_{\mathrm{K}}(k)$ only takes the discrete values $\pm 1$. The function $g\bm{(}v_{{\mathrm{K}},\star}(k),v_{{\mathrm{K}},\star}(k)\bm{)}$ thus becomes independent of the integration variable $k$ and is factored out, leaving only the integral of $P_{{\mathrm{K}},\star}^{(n)}(k)$. 
This yields:
\begin{align*}
\lim_{b_2 \to 0} \int_{\Sigma(\hat{\bm{v}})} g({\bm{v}})
\rho_{(a, b)}^{(1)}(\bm{v})d\bm{v}
& = \sum_{v = \pm 1} g(v, v) C_{v} \\
& = \int_{-1}^1 g(v,v)\rho_1(v) dv,
\end{align*}
where $\rho_1(v)$ is defined in \eqref{eq:1D(1)limitrepr} and the constants are $C_{\pm 1} = \left\|\Psi_0^{(\mp)}\right\|_{\ell^2(\mathbb{Z}^2)}^2$.

\bibliography{main}
\end{document}